# Formation of a simple cubic antiferromagnet through charge ordering in a double Dirac material

Tanya Berry[1,2*†‡], Vincent C. Morano[2*‡], Thomas Halloran[2†], Xin Zhang[3†], Tyler J. Slade[4,5], Aashish Sapkota[4,5], Sergey L. Bud'ko[4,5], Weiwei Xie[6†], Dominic H. Ryan[7], Zhijun Xu[8], Yang Zhao[8,9], Jeffrey W. Lynn[8], Tom Fennell[10], Paul C. Canfield[4,5], Collin L. Broholm[2,11], and Tyrel M. McQueen[1,2,11]

1. Department of Chemistry, The Johns Hopkins University, Baltimore, Maryland 21218, USA; 2. Institute for Quantum Matter and William H. Miller III Department of Physics and Astronomy, The Johns Hopkins University, Baltimore, Maryland 21218, USA; 3. Department of Chemical and Biomolecular Engineering, The Johns Hopkins University, Baltimore, Maryland 21218, USA; 4. Ames Laboratory, U.S. Department of Energy, Iowa State University, Ames, IA 50011, USA; 5. Department of Physics and Astronomy, Iowa State University, Ames, IA 50011, USA; 6. Department of Chemistry and Chemical Biology, Rutgers University, Piscataway, NJ 08854, USA; 7. Physics Department and Centre for the Physics of Materials, McGill University, 3600 University Street, Montreal, Quebec, H3A 2T8, Canada; 8. NIST Center for Neutron Research, National Institute of Standards and Technology, Gaithersburg, Maryland 20899, USA; 9. Department of Materials Science and Engineering, University of Maryland, College Park, Maryland 20742, USA; 10. Laboratory for Neutron Scattering and Imaging, Paul Scherrer Institut, 5232 Villigen PSI, Switzerland; 11. Department of Materials Science and Engineering, The Johns Hopkins University, Baltimore, Maryland 21218, USA

**ABSTRACT:** The appearance of spontaneous charge order in chemical systems is often associated with the emergence of novel, and useful, properties. Here we show through single crystal diffraction that the Eu ions in the mixed valent metal $EuPd_3S_4$ undergo long-range charge ordering at $T_{CO}$ = 340 K resulting in simple cubic lattices of $Eu^{2+}$ ($J = 7/2$) and $Eu^{3+}$ ($J = 0$) ions. As only one of the two sublattices has a non-magnetic ground state, the charge order results in the emergence of remarkably simple G-type antiferromagnetic order at $T_N$ = 2.85(6) K, observed in magnetization, specific heat, and neutron diffraction. Application of a 0.3 T field is sufficient to induce a spin flop transition to a magnetically polarized, but still charge ordered, state. Density functional theory calculations show that this charge order also modifies the electronic degeneracies present in the material: without charge order, EuPd$_3$S$_4$ is an example of a double Dirac material containing 8-fold degenerate electronic states, greater than the maximum degeneracy of six possible in molecular systems. The symmetry reduction from charge order transmutes 8-fold double Dirac states into 4-fold Dirac states, a degeneracy that can be preserved even in the presence of the magnetic order. Our results show not only how charge order can be used to produce interesting magnetic lattices, but also how it can be used to engineer controlled degeneracies in electronic states.

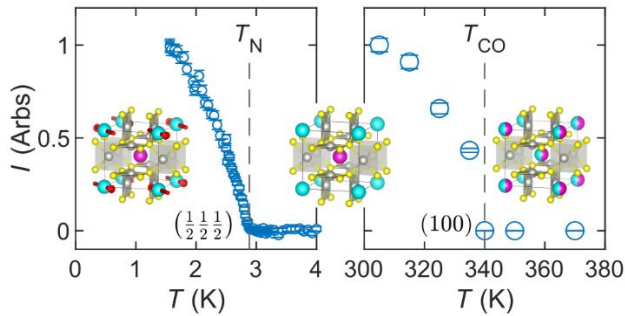

INTRODUCTION

Charge order has been observed to profoundly modify the properties of materials.[1–3] One property of both technological and fundamental importance is magnetic order. Despite the variety of magnetic structures observed in rare earth-based compounds, certain simple magnetic lattices are rarely encountered. It is important to identify techniques by which chemists may modify magnetic order to realize these simple but desirable structures. Palladium bronzes, of the general formula LnPd$_3$S$_4$ (Ln = rare-earth), consist of a three-dimensional network of corner sharing PdS$_4$ poly-anions with 8-fold sulfur coordinated cavities where Ln ions reside on a body-centered-cubic (bcc) lattice. For EuPd$_3$S$_4$, Mössbauer measurements[4,5] suggest a 1:1 ratio of Eu$^{2+}$ and Eu$^{3+}$, opening the possibility of two distinct sites for Eu$^{2+}$ and Eu$^{3+}$ and the formation of a Heisenberg magnet on the Eu$^{2+}$ lattice defined by charge ordering.

Further, interest has grown in materials that exhibit an interplay between degeneracy and interactions.[6–10] The highest known electronic degeneracy compatible with the 32 crystallographic point groups is six, i.e., t$_{2g}$ orbital states in an octahedral crystal field system. Recently, it was recognized[6,7] that nonsymmorphic symmetries, in particular the combination of a fractional translation and point symmetry operation, can raise the degeneracy at suitable high symmetry points of the Brillouin Zone to twelve (without spin-orbit coupling), or eight (with spin-orbit coupling). The nonsymmorphic $Pm\bar{3}n$ space group[11,12] of LaPd$_3$S$_4$ has been proposed to harbor 8-fold degenerate double Dirac states when the Ln cation is non-magnetic (e.g. La$^{3+}$).[7,13] With EuPd$_3$S$_4$ reported to contain partial Eu$^{2+}$,[4,5,14,15] the change in charge transfer to the [Pd$_3$S$_4$] framework should change the behavior and nature of the 8-fold degenerate points. The intermediate Eu valence provides additional richness with the possibility of 2-fold degenerate Weyl points depending on the magnetic order.

Using X-ray and neutron diffraction, we show that Eu$^{2+}$/Eu$^{3+}$ charge ordering indeed occurs and study the resulting Eu$^{2+}$ based simple cubic Heisenberg antiferromagnetism. We identify the 8-fold degeneracy in the band structure with density functional theory and examine how the loss of symmetry with both charge and magnetic order modifies this degeneracy.

RESULTS

Charge Ordering

We discovered the charge order in high resolution single crystal X-ray diffraction data acquired at $T$ = 213 K. While a reasonable refinement is obtained with a mixed Eu$^{2+}$/Eu$^{3+}$ site in space group $Pm\bar{3}n$, inspection of precession images revealed weak forbidden reflections at ($H$00) ($H$ odd) with an average of 5.5$\sigma$ ($\sigma$ = one standard deviation) discrepancy, and ($HHL$) ($L$ odd) positions with average 6.8$\sigma$ deviation, implying breaking of the n-glide symmetry. A more satisfactory refinement was obtained using $Pm\bar{3}$ (Table 1, S1). While our X-ray diffraction data are not directly sensitive to the 1e- difference between Eu$^{3+}$ and Eu$^{2+}$, the charge order is detectable due to the concomitant displacements of S anions, which yields distinct Eu-S bond lengths $d_{short}$ = 2.84604(5) Å and $d_{long}$ = 2.93528(5) Å. The corresponding bond valence sums, using $d_0$ = 2.48 Å are 2.8(1) and 2.2(1), implying formal valences of Eu$^{3+}$ and Eu$^{2+}$, respectively.

| | $T \leq 2.85(6)$ K | $2.85(6)$ K $\leq T \leq 340$ K | $T \geq 340$ K |
|---|---|---|---|
| Inversion symmetry | ✓ | ✓ | ✓ |
| Time reversal symmetry | ✗ | ✓ | ✓ |
| Nonsymmorphic symmetry | ✗ | ✗ | ✓ |
| Space group | $Pm\bar{3}$ | $Pm\bar{3}$ | $Pm\bar{3}n$ |
| Charge order | ✓ ($Eu^{2+}$ and $Eu^{3+}$, 1:1) | ✓ ($Eu^{2+}$ and $Eu^{3+}$, 1:1) | ✗ |
| Magnetism | ($\pi\pi\pi$) Antiferromagnetic | Paramagnetic | Paramagnetic |
| Band crossing | R point (2-/4-fold) | R point (4-fold) | R point (8-fold) |
| Structure $Eu^{2+}$, $Eu^{3+}$, Pd and S | 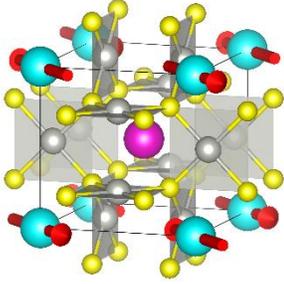 | 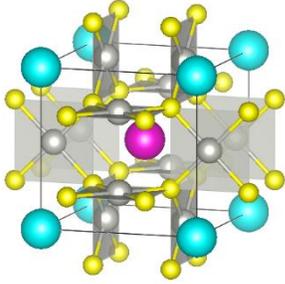 | 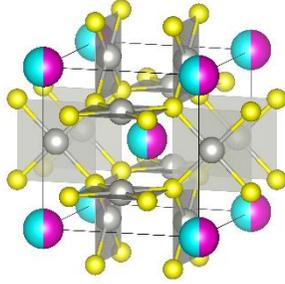 |

**Table 1. Overview of phase transitions in $EuPd_3S_4$.**[1]

The charge order and $Pm\bar{3}$ symmetry was confirmed by measurements at $T$ = 301(2) K on a different crystal and X-ray diffractometer. Figure 1a and Table S2 document the presence of the {100}-type reflections, which is forbidden in $Pm\bar{3}n$. Measurements on warming demonstrate that the (100) Bragg reflection disappears by $T_{CO}$ = 340 K (Figure 1b), which indicates restoration of the n-glide (refined structure at $T$ = 370(2) K, Table S3 and Figure S1). The critical temperature is consistent with extrapolation of the temperature of mixed valency behavior in the $La_{1-x}Eu_xPd_3S_4$ series to x = 1.0 (pure $EuPd_3S_4$).[5]

---

[1] Above $T$ = 340 K, the $Eu^{2+}/Eu^{3+}$ valence fluctuates, and $EuPd_3S_4$ adopts a $Pm\bar{3}n$ structure with inversion, time reversal, and a non-symmorphic n-glide that produce double Dirac fermions at the R point. The emergence of charge order below $T$ = 340 K removes the n-glide and results in gapping out of the double Dirac point to yield a pair of 4-fold degenerate states. This charge order produces a simple cubic magnetic lattice upon which G-type AFM order forms below $T$ = 2.85(6) K. Depending on the spin direction in the ground state, the 4-fold degeneracies are preserved or split into pairs of 2-fold degenerate states. Crystal structures were generated in VESTA[44] using a modified .cif from [12].

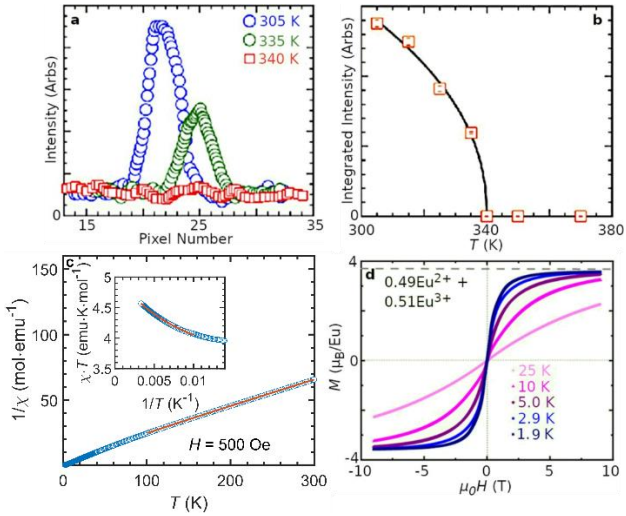

Figure 1. Charge and magnetic order develop in different temperature regimes in EuPd$_3$S$_4$. (a) Representative X-ray line scans of the $Pm\bar{3}$ allowed, $Pm\bar{3}n$ forbidden (100) Bragg reflection as a function of temperature showing its complete suppression on warming beyond $T_{CO}$ = 340 K. (b) Integrated intensities from fitting the line scans with a Gaussian lineshape indicates a phase transition at $T_{CO}$ = 340 K. The black line is to guide the eye. (c) Temperature-dependent magnetic susceptibility of a 74.9(1) mg powder sample under a $\mu_0 H$ = 0.05 T applied field upon warming. Note that uncertainty in the measured mass gives a 1.3% systematic uncertainty. The model includes the first nontrivial term of the high-$T$ expansion for the Eu$^{2+}$ susceptibility and the Eu$^{3+}$ Van Vleck susceptibility with $\lambda$ = 480 K.[16,17] Fitting from $T$ = 300 K to $T$ = 100 K while taking the Eu$^{2+}$:Eu$^{3+}$ ratio to be 1:1 yields $\theta_{CW} = -5.29(2)$ K and $\mu_{eff} = 7.95(1)$ $\mu_B$ for $\chi_r^2$ = 0.0004. We note that the high-$T$ data approach $T_{CO}$ = 340 K, so that charge fluctuations may play a role in this regime. The extracted $\theta_{CW}$ is a function of the fitting range with fits down to 125 K giving $\theta_{CW} = -5.72(1)$ K and 75 K giving $\theta_{CW} = -4.25(3)$ K. (d) $M(H)$ data show development of a field polarized state, saturating at a value consistent with a 1:1 mixture of Eu$^{2+}$ and Eu$^{3+}$. The field derivative at $T$ = 1.9 K (Figure S19) shows the AFM to canted AFM phase line as in Figure 3f.

Neutron diffraction in the presence of a magnetic field directly distinguishes Eu$^{2+}$ and Eu$^{3+}$ because magnetized Eu$^{2+}$ ($L$ = 0, $J$ = $S$ = 7/2) has a different scattering cross section compared to Van Vleck ion Eu$^{3+}$ ($J$ = 0). Indeed, we did detect the (100) Bragg peak by neutron diffraction upon magnetizing EuPd$_3$S$_4$ and the full temperature and field dependence of the intensities of both the (001) and (002) reflections are well-fit to squared Brillouin functions in the charge ordered magnetized paramagnetic phase (Figure S2-S3). This offers direct evidence that Eu$^{2+}$ forms a simple cubic lattice in the charge ordered state.

Magnetic Moments and Their Interactions

Charge ordering has profound impacts on the magnetism of EuPd$_3$S$_4$: It changes the magnetic lattice from bcc to simple cubic (sc) and eliminates magnetic bcc nearest neighbors. The eventual magnetic phase transition occurs at just 1% of the charge ordering temperature. For temperatures below 3 K, a field of $\mu_0 H$ = 3 T is sufficient to fully magnetize EuPd$_3$S$_4$ (Figure 1d). The resulting saturation magnetization of 3.7 $\mu_B$ is as expected for a 1:1 mixture of Eu$^{2+}$ and Eu$^{3+}$. To isolate the 4f Eu contribution to the magnetic susceptibility we measured and subtracted $\chi(T)$ for LaPd$_3$S$_4$. We plot the resulting molar susceptibility of EuPd$_3$S$_4$ in Figure 1c.

Heat capacity measurements versus field and temperature were employed to determine the overall magnetic entropy and the magnetic phase diagram. Using the heat capacity of LaPd$_3$S$_4$ as a non-magnetic reference,[13] the magnetic heat capacity and the magnetic entropy were determined and are displayed versus $T$ in Figure S6. The zero-field magnetic entropy integrates to $|\Delta S|$ = 8.0(5) J/mol-f.u/K, which is close to the value expected for 50% Eu$^{2+}$ ($J$ = 7/2), $0.5R\ln(8)$ = 8.6 J/mol-f.u./K. A lambda-like anomaly consistent with a second order phase transition is observed at $T_N$ = 2.85(6) K (Figure 2a). Analysis of this peak (Figure S8-S10) yields a critical exponent $\alpha = \alpha' \in [-0.3, -0.02]$. For reference, the exponents for the 3D Heisenberg and mean field classes are $\alpha = \alpha' = -0.12(1)$ and $\alpha = \alpha' = 0$, respectively.[18] Upon applying a magnetic field along $(1\bar{1}0)$, the peak progressively shifts to lower temperatures until for $\mu_0 H$ = 3 T, there is no longer a phase transition. The small upturn in $C(T)$ for $T < 0.3$ K is attributed to the polarization of the Eu$^{2+}$ nuclear spin system in the hyperfine enhanced field generated by the electronic sublattice polarization (Figure S7 and Table S4). Just above $T_N$, for $T$ = 3.15(7) K the heat capacity is approximately independent of applied field while $\mu_0 H < 3$ T.

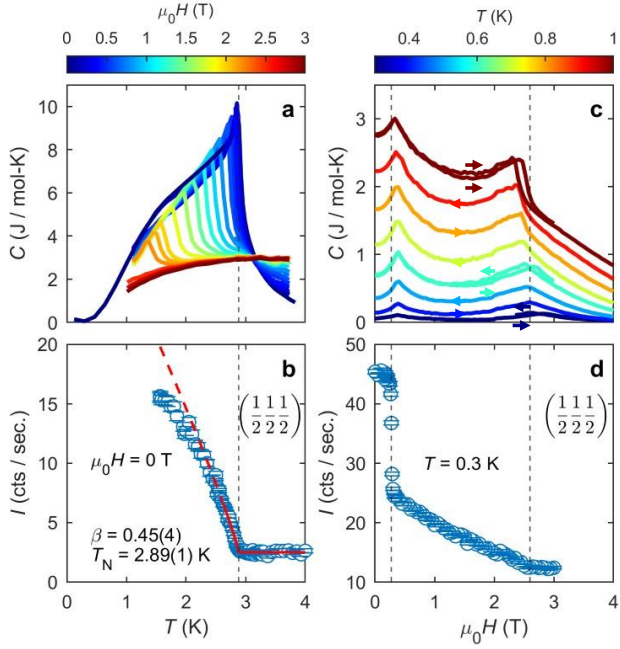

Figure 2. Specific heat and neutron diffraction measurements indicate a magnetic field tunable second order phase transition. (a) Temperature-dependent heat capacity scans at fixed fields from $\mu_0 H$ = 0 T to 3 T from two experiments on the same EuPd$_3$S$_4$ single crystal with the field applied along the (122) direction. Fields include 0.0 T, 0.1 T, 0.2 T, 0.23 T, 0.3 T, 0.37 T, and 0.4 T to 0.8 T in 0.1 T increments, 1.0 T to 2.0 T in 0.2 T increments, 2.1 T, 2.2 T, 2.5 T, 2.75 T, and 3.0 T. (b) Order parameter measurement of the (½,½,½) peak from neutron diffraction at $\mu_0 H$ = 0 T, indicating $T_N$ = 2.89(1) K. The solid line is a fit of the intensity to a power law (Figure S11), while the dashed line shows this model's behavior well beyond the critical regime. (c) Field-dependent heat capacity scans at fixed temperatures from $T$ = 0.3 K to 1 K in 0.1 K increments. Measurements above 1 K show the peak at lower field becomes a shoulder (Figure S15). Arrows indicate the direction of each field scan. (d) Field scan of the $\left(\frac{1}{2}\frac{1}{2}\frac{1}{2}\right)$ peak intensity at $T$ = 0.3 K. The same two transitions from (c) are observed at low and high field. The peak vanishes beyond the upper transition. The lower field step is attributable to the transition from a simple multidomain state to a smaller number of domains with staggered magnetization perpendicular to the applied field and the moments canted along the applied field. The magnetic structure becomes entirely polarized along the field beyond the upper transition.

Magnetic Order

To determine the ordered magnetic structure, we use magnetic neutron diffraction. EuPd$_3$S$_4$ poses challenges for neutron scattering as natural Eu absorbs neutrons, with an absorption coefficient $\mu_a$ = 2.60 mm$^{-1}$ (SI) at the $\lambda$ = 1.5 Å wavelength employed in our measurements.[19] Additionally, cubic symmetry implies that magnetic neutron diffraction from a multidomain sample is independent of the moment direction within each domain. Nonetheless, we were able to establish key aspects of the magnetic order and the magnetic phase diagram through neutron diffraction from mm-scale EuPd$_3$S$_4$ single crystals with the natural isotopic abundance. Contrary to expectations from theory[14] and experiments in related materials,[20] scans through reciprocal space at $T$ = 1.6 K and $\mu_0 H$ = 0 T show that Bragg diffraction at (001) is induced when applying a field (Figure 3a, discussed further below). The zero-field magnetic ordering is instead associated with magnetic Bragg diffraction at $\left(\frac{1}{2}\frac{1}{2}\frac{1}{2}\right)$ (Figure 3b). This indicates a doubling of the unit cell along the a, b, and c axes. Consistent with heat capacity and magnetization data, Figure 2b shows the $\left(\frac{1}{2}\frac{1}{2}\frac{1}{2}\right)$ peak onsets below $T_N$ = 2.89(1) K (see Figure S11). As a function of field applied along (1$\bar{1}$0), we find a sharp decrease in the $\left(\frac{1}{2}\frac{1}{2}\frac{1}{2}\right)$ peak intensity near $\mu_0 H$ = 0.3 T followed by a linear decrease that culminates in the complete disappearance of this AFM diffraction for $\mu_0 H$ = 3 T (Figure 2d). Both anomalies have counterparts in constant temperature field sweeps of the heat capacity (Figure 2c). We infer that the lowest field transition is associated with alignment of the staggered magnetization perpendicular to the applied field. The subsequent decreasing intensity indicates spin canting that gradually reduces the staggered magnetization until, at $\mu_0 H$ = 3 T, EuPd$_3$S$_4$ is fully magnetized.

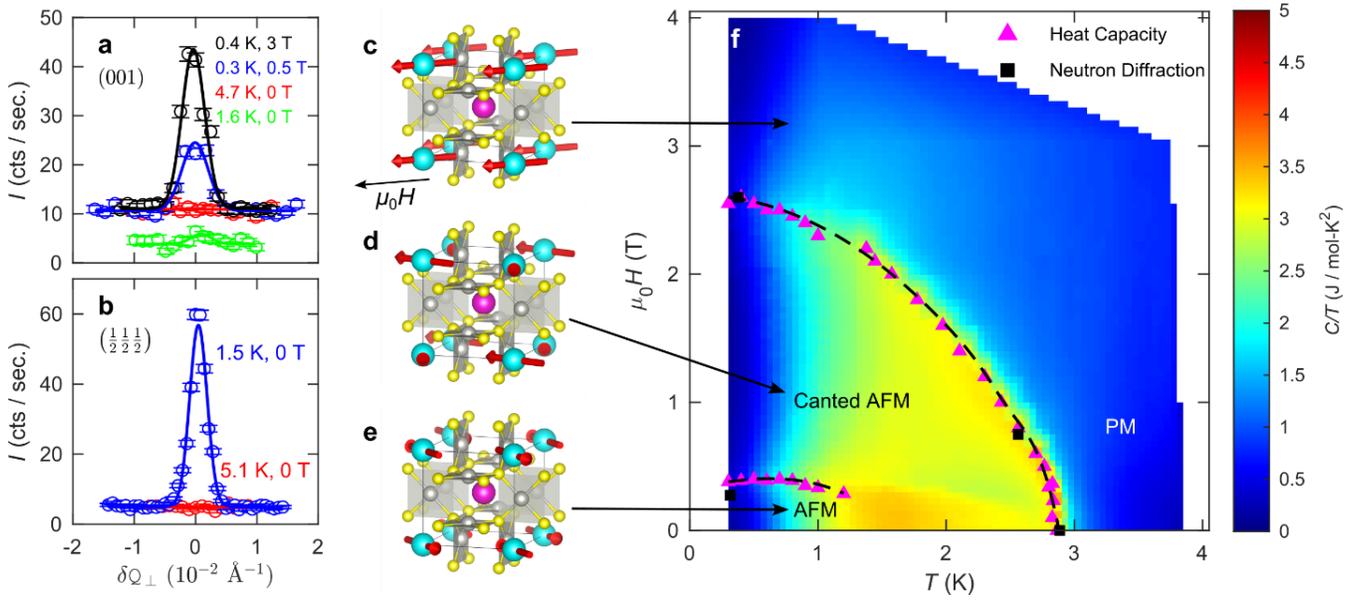

Figure 3. Field-dependent magnetic phase diagram. (a) Rocking scans with δQ⊥ perpendicular to the (001) direction of wavevector transfer, lying in the (*HHL*) plane. Green points were collected without a cryo-magnet so that the background is lower. The intensity has been scaled to the intensity of the (112) peak measured with a 70 mm Orange cryo-magnet so peak intensity data can be directly compared for the two configurations employed. The (001) reflection has zero nuclear Bragg intensity at *T* = 4.7 K because the nuclei of $Eu^{2+}$ and $Eu^{3+}$ are equivalent. (001) remains absent below $T_N$ in zero magnetic field. As the applied field along (1-10) increases and the $Eu^{2+}$ sublattice becomes magnetized, however, the (001) peak appears. (b) Rocking scans with δQ⊥ perpendicular to (½,½,½) and within the (*HHL*) reciprocal lattice plane, showing the development of AFM order below $T_N$. Gaussian fits are shown in frames (a-b). (c) The field polarized magnetic state in which all magnetic $Eu^{2+}$ atoms are aligned along the applied (1-10) field direction. (d) The intermediate canted AFM state in which the $Eu^{2+}$ spins cant along the applied field. (e) The multidomain (½,½,½) AFM state in the absence of an applied field. Due to domain averaging, the moment directions in the zero field state cannot be determined by neutron diffraction (SI). Because a small (1-10) field forces the staggered magnetization along (110), we show the moments oriented along (110). (f) Resultant magnetic phase diagram for $EuPd_3S_4$. The points indicating phase transitions were determined by inspection of anomalies in the field and temperature dependence of the underlying specific heat and neutron scattering data.

For magnetic structure refinement, we acquired sets of neutron diffraction peaks in the PM, AFM, and canted states (Figure S12, Table S5). Below $T_N$, $\boldsymbol{k}_c = \left(\frac{1}{2}\frac{1}{2}\frac{1}{2}\right)$ AFM peaks indicate the zero-field low temperature state (Figure 3e). Of note is the emergence of Bragg diffraction at (001) upon application of a field (Figure 3a), which implies a field induced difference in the neutron scattering amplitude between the corner and center Eu sites of the bcc lattice. This is consistent with the observation of $Eu^{2+}/Eu^{3+}$ charge ordering in the X-ray data. The resolution limited nature of the field induced (001) peak yields a lower bound of 490(20) Å on the correlation length for the charge ordered state (Figure S13-S14). Refinements of the $\boldsymbol{k}_c = \left(\frac{1}{2}\frac{1}{2}\frac{1}{2}\right)$ diffraction data acquired under a 0.5 T field applied along $(1\bar{1}0)$ with simple cubic charge order and spherical absorption correction indicates the staggered magnetization is oriented along [110] rather than [001] direction (Figure 3d). At $\mu_0 H = 3$ T, the $\boldsymbol{k}_c = \left(\frac{1}{2}\frac{1}{2}\frac{1}{2}\right)$ peaks disappear indicating a uniform magnetization oriented entirely along the applied field (Figure 3c). Together with additional field-dependent heat capacity sweeps (Figure S15), these data establish the magnetic phase diagram for $EuPd_3S_4$ (Figure 3f).

Electronic Structure

Density functional theory (DFT) calculations (Figure 4, S16) including spin-orbit coupling for the high temperature structure (Figure 4c) show protected 8-fold crossings at the $R\left(\frac{1}{2}\frac{1}{2}\frac{1}{2}\right)$ point in the Brillouin Zone (BZ). As there is only a single crystallographic site, DFT cannot capture distinct Eu valences, and the total charge transfer to the $[Pd_3S_4]$ framework is similar to that found for $La^{3+}$.[13] A symmetry analysis shows that the 8-fold Fermion just above $E_F$ transforms as the $\Gamma_6^- + \Gamma_7^- + \Gamma_6^+ + \Gamma_7^+$ irreducible representations, with the states of opposite parity constrained to be equivalent due to the non-symmorphic n-glide. A DFT calculation including charge order (Figure 4b), however, can capture distinct Eu valences since there are now two distinct Eu sites. This results in less charge transfer to the [Pd$_3$S$_4$] poly-anion, indicated by a lowering of $E_F$ relative to rigid bands.

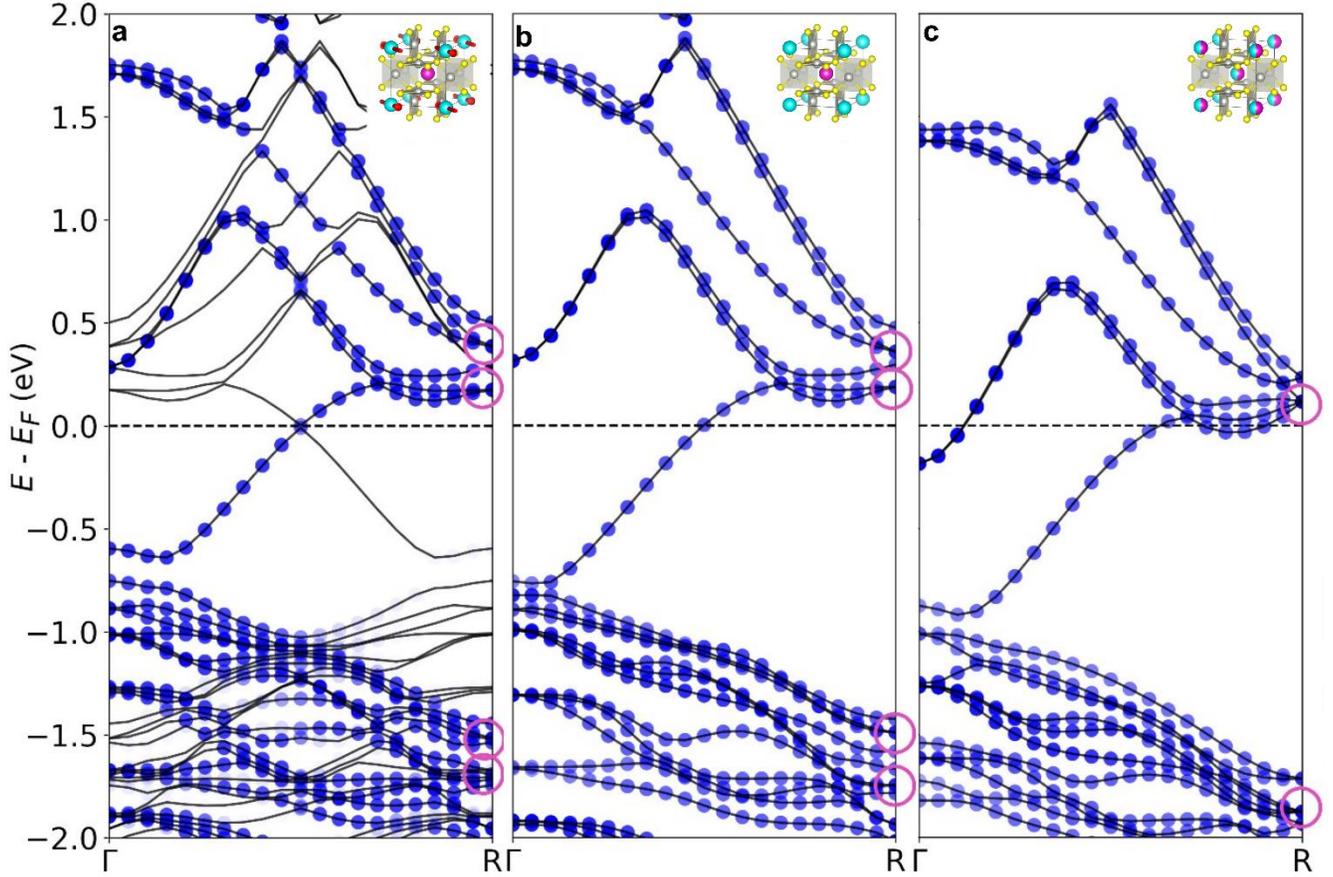

Figure 4. Spin-Orbit Density Functional Theory band structures from Γ (0,0,0) to R (½,½,½) showing charge and magnetic order transmuting an 8-fold fermion (a) Unfolded band structure from the G-type simple cubic AFM magnetic state below $T_N$ = 2.85(6) K showing retention of the 4-fold fermions when the magnetic moments lie along an axis with $Fmmm'$ magnetic symmetry. Solid lines show bands as computed, blue dots indicate band weight in the unfolded BZ. (b) Below 340 K, Eu$^{2+}$/Eu$^{3+}$ charge order, corresponding to $Pm\bar{3}$ crystal symmetry, splits the 8-fold fermions into pairs of 4-fold fermions, circled. (c) The crystal structure of EuPd$_3$S$_4$ at $T$ = 370(2) K corresponds to $Pm\bar{3}n$ crystal symmetry, with a single crystallographically unique Eu site; the n glide plane protects 8-fold degenerate states at the R point, circled.

## DISCUSSION

In the low T regime where charge is localized on the simple cubic lattice, the magnetism of EuPd$_3$S$_4$ should be well described by a spin Hamiltonian of the form $\hat{H} = -\frac{1}{2}\sum_{R,R'} J(R - R')S(R) \cdot S(R')$ and the random phase approximation (RPA) should provide a reasonable approximation to the generalized wave vector dependent susceptibility: $\chi_\mathbf{k}(T) = \frac{\chi_0(T)}{1-\chi_0(T)J(\mathbf{k})}$. Here $J(\mathbf{k})$ is the lattice Fourier transform of the exchange constants $J(\mathbf{R})$ and $\mathbf{R}$, $\mathbf{R}'$ denote simple cubic Bravais sites. In the field-polarized state, the spin wave dispersion relation reads $E_\mathbf{k} = S[J(\mathbf{k} = \mathbf{0}) - J(\mathbf{k})] + g\mu_B B$.[21] From $\chi_\mathbf{k}(T)$ and $E_\mathbf{k}$ we obtain expressions for three collective energy scales $k_B \Theta_{CW} = \frac{S(S+1)}{3}J(\mathbf{k} = \mathbf{0})$, $k_B T_N = \frac{S(S+1)}{3}J(\mathbf{k}_c)$, and $g\mu_B B_c(T = 0) = S[J(\mathbf{k}_c) - J(\mathbf{k} = \mathbf{0})]$. Combining these we obtain $\Theta_{\mathrm{CW}} = T_{\mathrm{N}} - \frac{(S+1)g\mu_{\mathrm{B}}}{3k_{\mathrm{B}}}B_{\mathrm{c}} = -2.6(4)$ K. We obtain $\Theta_{\mathrm{CW}} = -5.29(2)$ K from our susceptibility measurements (Figure 1c) while reference [4] reports $\Theta_{\mathrm{CW}} = 1.0$ K. Considering only nearest and next nearest neighbor interactions (though longer range RKKY interactions may be significant) we find $J_1 = -\frac{g\mu_{\mathrm{B}} B_{\mathrm{c}}}{12 k_{\mathrm{B}} S} = -86(6)$ mK and $J_2 = \frac{T_{\mathrm{N}}}{4S(S+1)} - \frac{g\mu_{\mathrm{B}} B_{\mathrm{c}}}{24 k_{\mathrm{B}} S} = 2(3)$ mK, favoring $\boldsymbol{k}_{\mathrm{c}} = \left(\frac{1}{2}\frac{1}{2}\frac{1}{2}\right)$ type AFM order (Figure 3e).

Let us now consider the interplay between the collective charge and spin order in EuPd$_3$S$_4$ on the electronic band structure. The charge order breaks the n-glide and lifts the degeneracy of states of opposite parity at R, resulting in 4-fold Fermions $\Gamma_6^- + \Gamma_7^-$ and $\Gamma_6^+ + \Gamma_7^+$. This demonstrates an alternative avenue by which 8-fold degeneracies can be lifted.[22,23] The AFM order for $T < T_{\mathrm{N}} = 2.85(6)$ K, however, does not necessarily further reduce the electronic degeneracy (Figure 4a). With G-type AFM and spins along a crystallographic symmetry axis, the magnetic group symmetry is *Fmmm'*. Unfolding back to the primitive BZ shows that the 4-fold degeneracies at R remain, transforming as $2\Gamma_3^+ + 2\Gamma_4^+$ in the $C_{2h}$ double group. Given the lack of large rearrangements of the band structures, no exceptional transport properties are expected, and none are observed (Figure S17). However, considering the proximity of the 8-fold fermion to $E_{\mathrm{F}}$, future work should explore whether charge ordering might be understood as an instability of double Dirac states[24] as in Jahn-Teller distortions where the energy is lowered by breaking symmetries that protect degeneracies, More broadly, our results demonstrate an intricate interplay between charge and spin order that conspire to alter the electronic topology of EuPd$_3$S$_4$.

METHODS

Synthesis

EuPd$_3$S$_4$ single crystals were grown by two approaches, inspired by previous related syntheses.[13,25–28] In the first approach, "crystal A", a self-flux technique, stoichiometric quantities of Eu (ingot, Yeemeida Technology Co., LTD 99.995%), Pd (J&J materials, 99.9+%), and S (Alfa Aesar, 99.999+%) were sealed in evacuated quartz ampoules with total mass of 0.5 g. The ampoules were placed in a box furnace at a 45° angle (with the sample charge close to the interiors of the furnace) and heated to $T$ = 900°C at a rate of 80°C/hr for 24 hr and subsequently heated to $T$ = 1100°C at a rate of 80°C/hr for 14 days. The ampoules were removed at $T$ = 1100°C and were allowed to air cool and opened at room temperature. The obtained crystals were 1-3 mm in size. Phase purity and single crystal domain size were characterized using powder X-ray diffraction and Laue diffraction (Figure S18). In the second approach, "crystal B", a ternary solution using a two-step process described previously was used.[25] Elemental Eu (Ames Lab. 99.9%), Pd powder (Engelhard, 99+% purity), and S (Alpha-Aesar, 99.99%) were weighed according to a nominal composition of Eu$_5$Pd$_{57}$S$_{38}$ and loaded into 2 mL alumina Canfield Crucible sets, or CCS.[27] The crucible sets were flame sealed inside evacuated fused silica ampoules, and prior to sealing, the ampoules were back-filled with ~1/6 atm Ar gas. The ampoules were then heated in a box furnace to 450°C over 8 h and then to 1150°C in another 8 h, held at 1150°C for 6 h, and finally cooled to 1050°C over 6 h, upon which the excess liquid was decanted by inverting the ampoules into a centrifuge. In our previous work, we found that mixtures of Ln (rare-earth), Pd, and S generally do not form a clean, single phase, liquid below 1200°C, and this first step allows us to remove unwanted secondary phases and obtain a clean solution for crystal growth.[25] Upon cooling to room temperature, a second CCS was assembled, using the original "catch" crucible and the material within (that was captured during decanting) as the starting material. The second CCS was evacuated and sealed as described above, heated to 1075°C, and allowed to dwell for 6 h, during which the decant melted into a single-phase liquid. The furnace was finally slowly cooled to 875°C in ~150 h, where the excess liquid phase was again decanted. The ampoules and crucibles were opened to reveal large (up to 5-6 mm), well faceted crystals of EuPd$_3$S$_4$.

X-Ray Diffraction

Single crystal X-ray diffraction data at $T$ = 213 K on "crystal A" was collected on a BrukerNonius X8 Proteum (Mo K$_\alpha$ radiation) diffractometer equipped with an Oxford cryostream. Integration and scaling were performed using CrysalisPro (Version 1.171.39.29c, Rigaku OD, 2017). A multiscan absorption correction was applied using SADABS.[29] The structure was solved by direct methods, and successive interpretations of difference Fourier maps were followed by least-squares refinement using SHELX and WinGX.[30,31] For "crystal B", single crystal X-ray diffraction measurements were performed at $T$ = 370 K and $T$ = 301 K. The data were collected over a full sphere of reciprocal space with 0.5° scans in ω with an exposure time of 10 seconds per frame. All measurements were performed using a Bruker Apex II diffractometer with Mo radiation K$_{\alpha 1}$ ($\lambda$ = 0.71073 Å). The 2$\theta$ range was from 5° to 75°. The SMART software was used for data acquisition. The geometric Lorentz factor has been subtracted from the scattering intensities using the SAINT program. Numerical absorption corrections were accomplished with XPREP which is based on face-indexed absorption. With the SHELXTL package, the crystal structures were solved using direct methods and refined by full-matrix least-squares on $F^2$.[30,31] The disorder refinements on Eu, Pd, and S sites show no vacancies in either EuPd$_3$S$_4$ crystal to within instrumental precision (~3%). To acquire the temperature dependence of the forbidden (100) peak, measurements were done with a narrow $2\theta$ range around the peak and the temperature was controlled using the liquid nitrogen stream.

Magnetization and Specific Heat

The magnetic and specific heat measurements were performed with Quantum Design Magnetic and Physical Properties Measurement Systems (MPMS-7T, PPMS-9T and PPMS-14T), the latter equipped with a dilution refrigerator. $M(T)$ measurements from $T$ = 2 K to 300 K were performed using the MPMS-7T, with an applied field of $\mu_0 H$ = 0.05 T. A 74.9(1) mg powder sample was wrapped in plastic and mounted in a plastic drinking straw. Isothermal $M(H)$ measurements were performed at $T$ = 1.8 K, 2.9 K, 5 K, 10 K, and 25 K, using the ACMS option in DC mode. Magnetization measurements from $T$ = 2 K to 6 K were performed using the MPMS in DC mode with an applied field of $\mu_0 H$ = 2 mT. For heat capacity measurements under field, a thin 0.19(3) mg $EuPd_3S_4$ plate was aligned along (122) using Laue diffraction (Figure S17) and measured using the semi-adiabatic method with a 2% temperature rise. Field and temperature scans were performed in the range from $T$ = 0.3 K to 4 K and $\mu_0 H$ = 0 T to $\mu_0 H$ = 4 T using the DR option. Additional measurements on a 2.34(3) mg crystal with larger temperature rises were performed and analyzed with LongPulse to search for the latent heat at the magnetic phase transition (Figure S5).[32] The absolute scale for the specific heat data measured with the 0.19(3) mg sample was determined by comparison to the results for the 2.34(3) mg sample where the mass could be measured with higher relative accuracy.

Neutron Diffraction

Neutron diffraction was performed over the course of two experiments on the BT7 triple-axis spectrometer at the NIST Center for Neutron Research (NCNR)[33] and a third experiment on EIGER at PSI. A 30 mg $EuPd_3S_4$ single crystal was attached to an oxygen free copper sample holder with cadmium shielding and aligned with a horizontal ($HHL$) scattering plane by Laue diffraction.[34] For the first experiment, the sample was placed in an ILL-type cryostat. For the second experiment, the sample was placed in a 7 T cryo-magnet system (with the applied field vertical along (1-10)) and cooled in a ³He insert. In both cases the neutron energy was $E_i = E_f$ = 35 meV, collimation was open-25'-25'-120', and the monochromator and analyzer crystals were PG(002). Measurements included survey scans through the BZ, and measurements of rocking scans for accessible magnetic and nuclear Bragg points at base temperature, above $T_N$, at 0.5 T, and at 3 T. Temperature scans and field scans were acquired while monitoring the (½,½,½) peak intensity. Preliminary analysis was performed using DAVE[35] and MATLAB R2023a, with the refinement completed in FullProf[36] (Figure S12, Table S5). Magnetic propagation vector analysis was carried out using SARAh[37] and the ISOTROPY tools.[38-40] For our third neutron diffraction experiment, a 2.35 mg $EuPd_3S_4$ single crystal was attached to an Al sample holder using Cytop. The holder was wrapped in Cd shielding. The crystal was aligned for diffraction in the ($HHL$) scattering plane by Laue X-ray backscattering and loaded into an MA10 magnet on a dilution insert for neutron diffraction on EIGER at PSI. A field of 0.05 T was applied throughout the experiment to suppress Al superconductivity and ensure good thermal contact to the cold finger of the dilution refrigerator. $\boldsymbol{k}_c = \left(\frac{1}{2}\frac{1}{2}\frac{1}{2}\right)$ peaks appeared below $T_N$, while both (001) and (002) appeared upon application of field at $T$ = 0.1 K. The temperature dependence of both peaks was measured up to $T$ = 50 K in a $\mu_0 H$ = 8 T applied field. These results are presented in Figure S2-S3.

Band Structure Calculations

Density functional theory calculations for $EuPd_3S_4$, including band symmetry analysis, were performed using Quantum Espresso 7.0[41] compiled with the Intel compiler 2020.1 on the Advanced Research Computing at Hopkins Rockfish cluster. Kinetic energy cutoffs and k-point meshes were chosen to ensure convergence with respect to those parameter choices. For non-spin-orbit calculations, PAW pseudopotentials with the LDA functional, non-linear core correction, and a scalar relativistic treatment, obtained from quantum-espresso.org, were used, with a 75 Ryd kinetic energy cutoff and an 11x11x11 k-point mesh. For non-spin-orbit, non-magnetic calculations, the 4f electrons were in the pseudopotential core. For magnetic calculations, the $Eu^{2+}$ ions used PAW pseudopotentials with 4f electrons in the valence, from the Virtual Laboratory for Earth and Planetary Materials,[42] with $U$ = 10 eV. For spin-orbit calculations in $Pm\bar{3}n$, PAW pseudopotentials with the LDA functional, non-linear core correction, 4f electrons in the core, and a full relativistic treatment, obtained from quantum-espresso.org, were used, with a 120 Ryd kinetic energy cutoff and an 11x11x11 k-point mesh. For spin-orbit, non-magnetic calculations in $Pm\bar{3}$, the same conditions were used, except with custom Eu PAW, LDA, non-linear core correction, fully relativistic, pseudopotentials with 4f⁶ electrons in the core (for $Eu^{3+}$), and 4f⁷ electrons in the core (for $Eu^{2+}$), generated using the "atomic" code.[43] For spin-orbit, magnetic calculations in $Pm\bar{3}$ and $Fmmm'$, the $Eu^{2+}$ pseudopotential was replaced with a custom PAW, LDA, non-linear core correction, fully relativistic, pseudopotential, with 4f⁷ electrons in the valence. In magnetic calculations, moments were taken to lie along the crystallographic c axis. Calculations in the magnetic group $Fmmm'$ were performed with a crystallographic symmetry of $Fm\bar{3}$. Band unfolding was performed using a modified version of the code available at https://github.com/yw-choi/qe_unfolding.

## ASSOCIATED CONTENT

Supporting Information. Includes figures, tables, and text providing further insight but not essential to the understanding of the main text.

## AUTHOR INFORMATION

### Corresponding Authors

*Correspondence and material requests should be addressed to TBe and VCM at tberry@princeton.edu and vmorano1@jhu.edu, respectively.

### Present Addresses

†TBe: Department of Chemistry, Princeton University, Princeton, New Jersey 08544, USA; TH:
NIST Center for Neutron Research, National Institute of Standards and Technology, Gaithersburg, MD, USA; XZ: Department of Chemistry, Princeton University, Princeton, New Jersey 08544, USA; WX: Department of Chemistry and Chemical Biology, Rutgers University-New Brunswick, Piscataway, NJ, 08854, USA

### Author Contributions

‡These authors contributed equally. TBe grew single crystal via grain growth. TS and SLB grew single crystals via flux growth under the supervision of PCC. TBe, WX, and AS carried out the X-ray diffraction studies, with DHR helping reconcile results with literature Mössbauer measurements. VCM, ZX, YZ, JWL, and TF conducted the neutron studies under the supervision of CLB. TBe, TH, XZ, and VCM carried out the specific heat and magnetization measurements under the supervision of CLB and TMM. TMM carried out the first principles calculations. TBe carried out the resistivity measurements. All authors contributed to data interpretation and the writing and editing of the manuscript.

### Funding Sources

This research was conducted at the Institute for Quantum Matter, an Energy Frontier Research Center funded by the U.S. Department of Energy Office of Science, Basic Energy Sciences, under Award No. DE-SC001933. Calculations were performed using computational resources of the Maryland Advanced Research Computing Center and the Advanced Research Computing at Hopkins (ARCH) Rockfish cluster. The Rockfish cluster is supported by the National Science Foundation Award No. OAC-1920103. The MPMS was funded by the National Science Foundation, Division of Materials Research, Major Research Instrumentation Program, under Award No. 1828490. TBe acknowledges support from NSF-MRSEC through the Princeton Center for Complex Materials NSF-DMR-2011750. X. Zhang acknowledges support from the National Science Foundation Platform for the Accelerated Realization, Analysis and Discovery of Interface Materials (PARADIM) under Cooperative Agreement No. DMR-1539918. Work at Ames National Laboratory (TJS, AS, PCC) was supported by the U.S. Department of Energy, Office of Science, Basic Energy Sciences, Materials Sciences and Engineering Division. Ames National Laboratory is operated for the U.S. Department of Energy by Iowa State University under Contract No. DE-AC02-07CH11358. TJS and PCC were supported by the Center for Advancement of Topological Semimetals (CATS), an Energy Frontier Research Center funded by the U.S. Department of Energy Office of Science, Office of Basic Energy Sciences, through the Ames National Laboratory under its Contract No. DE-AC02-07CH11358 with Iowa State University. This work is based on neutron experiments performed at the NIST Center for Neutron Research and the Swiss spallation neutron source SINQ, Paul Scherrer Institute, Villigen, Switzerland. The identification of any commercial product or trade name does not imply endorsement or recommendation by the National Institute of Standards and Technology. CLB was supported by the Gordon and Betty Moore foundation EPIQS program under Grant No. GBMF9456.

### Notes

All data are available upon request from the authors. See Author Contributions for specific data sets. Neutron data from BT7 is publicly available at https://ncnr.nist.gov/ncnrdata/search.php under the date range from 2020-01-13 11:13 to 2023-04-20 15:48 and from 2020-03-04 15:43 to 2020-03-08 12:05. Please contact VCM regarding questions about the neutron data. All code is available upon request from the authors. See Author Contributions for specific code.

## ACKNOWLEDGEMENTS

TBe acknowledges Maxime Siegler's assistance with X-ray data collection. TBe and TMM appreciate useful conversations with Nicodemos Varnava and Qun Yang on the DFT calculations. TBe appreciates help with high field transport measurements from Tobias Förster and the support of the HLD at HZDR, a member of the European Magnetic Field Laboratory (EMFL).

# Supplementary Information for: Formation of a simple cubic antiferromagnet through charge ordering in a double Dirac material

Tanya Berry[1,2†‡], Vincent C. Morano[2‡], Thomas Halloran[2†], Xin Zhang[3†], Tyler J. Slade[4,5], Aashish Sapkota[4,5], Sergey L. Bud'ko[4,5], Weiwei Xie[6†], Dominic H. Ryan[7], Zhijun Xu[8], Yang Zhao[8,9], Jeffrey W. Lynn[8], Tom Fennell[10], Paul C. Canfield[4,5], Collin L. Broholm[2,11], and Tyrel M. McQueen[1,2,11]

1. Department of Chemistry, The Johns Hopkins University, Baltimore, Maryland 21218, USA; 2. Institute for Quantum Matter and William H. Miller III Department of Physics and Astronomy, The Johns Hopkins University, Baltimore, Maryland 21218, USA; 3. Department of Chemical and Biomolecular Engineering, The Johns Hopkins University, Baltimore, Maryland 21218, USA; 4. Ames Laboratory, U.S. Department of Energy, Iowa State University, Ames, IA 50011, USA; 5. Department of Physics and Astronomy, Iowa State University, Ames, IA 50011, USA; 6. Department of Chemistry and Chemical Biology, Rutgers University, Piscataway, NJ 08854, USA; 7. Physics Department and Centre for the Physics of Materials, McGill University, 3600 University Street, Montreal, Quebec, H3A 2T8, Canada; 8. NIST Center for Neutron Research, National Institute of Standards and Technology, Gaithersburg, Maryland 20899, USA; 9. Department of Materials Science and Engineering, University of Maryland, College Park, Maryland 20742, USA; 10. Laboratory for Neutron Scattering and Imaging, Paul Scherrer Institut, 5232 Villigen PSI, Switzerland; 11. Department of Materials Science and Engineering, The Johns Hopkins University, Baltimore, Maryland 21218, USA

Contents

Supplementary Figures

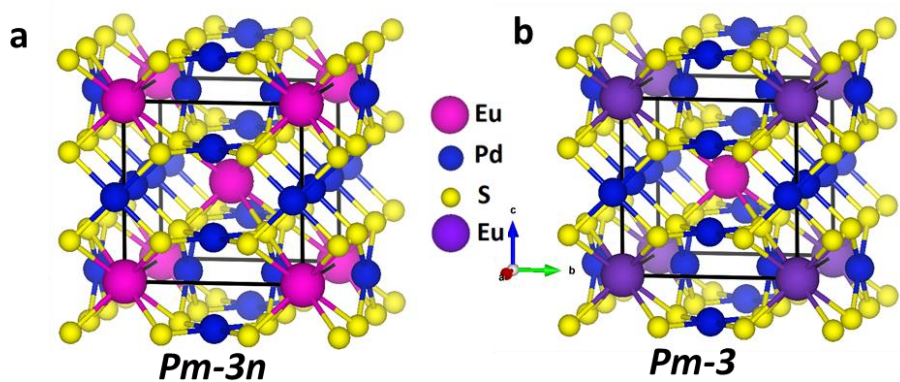

**Figure S1.** Crystal structures of EuPd$_3$S$_4$ in the charge ordered and charge disordered states. (a) Crystal structure at $T$ = 370 K corresponding to $Pm\bar{3}n$ crystal symmetry. (b) Crystal structure at $T$ = 301 K corresponding to $Pm\bar{3}$ crystal symmetry. In (b) sulfur has moved along body diagonal directions 0.045 Å closer to Eu1, which indicates this Eu1 is trivalent and Eu2 is divalent. Please note that the structures look identical because $Pm\bar{3}$ is a maximal non-isomorphic subgroup of space group $Pm\bar{3}n$ and the S-displacements are just 0.7% of the lattice parameter $a$.

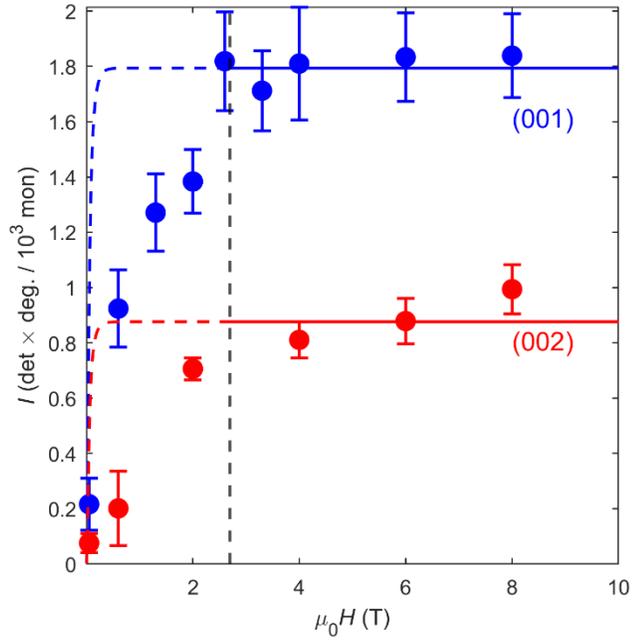

**Figure S2.** Single crystal neutron diffraction field scans with field perpendicular to the (HHL) scattering plane. Integrated intensity of the (001) and (002) Bragg peaks are shown as a function of applied field measured at $T$ = 0.1 K. For the $Pm\bar{3}$ simple cubic charge ordered system (Table 1) both peaks probe the uniform magnetization squared versus field. However, for the $Pm\bar{3}n$ space group (001) probes a staggered magnetization squared while (002) still probes the squared uniform magnetization. Up to high fields, at which we expect the moment to saturate, we find an indistinguishable field dependent integrated intensity for the two Bragg peaks. The dashed line fits are to a Brillouin function squared. The dashed vertical line at $\mu_0H$ = 2.7 T indicates the lower fitting limit below which we expect the sample to be antiferromagnetically ordered based on our $T$ = 0.3 K field scans from BT7 (Figure 2d). Data from EIGER at PSI.

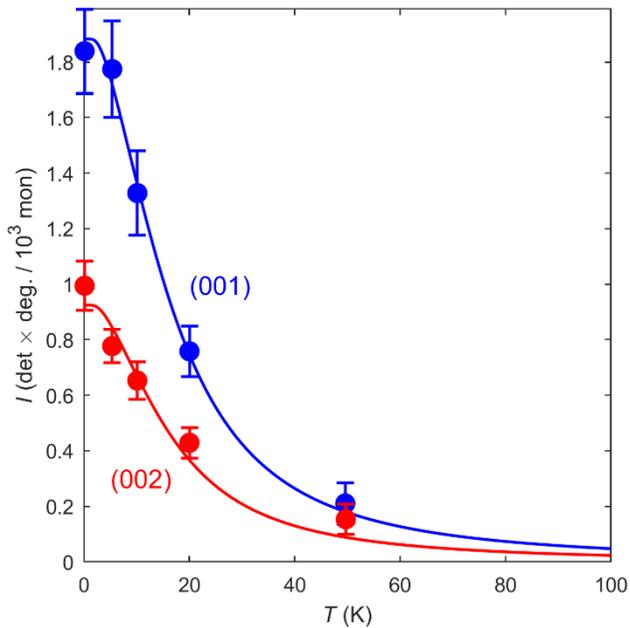

**Figure S3.** Single crystal neutron diffraction temperature scans. Peaks are scanned as a function of temperature at $\mu_0H$ = 8 T, with the field perpendicular to (HHL). Up to the highest temperatures, we see no significant deviation from fits to a squared Brillouin function. In the absence of charge ordering, high field magnetic Bragg diffraction at (001) should vanish. Consistent with Figure 1, there appears to be no changes in the charge order in the temperature regime probed here. Data from EIGER at PSI.

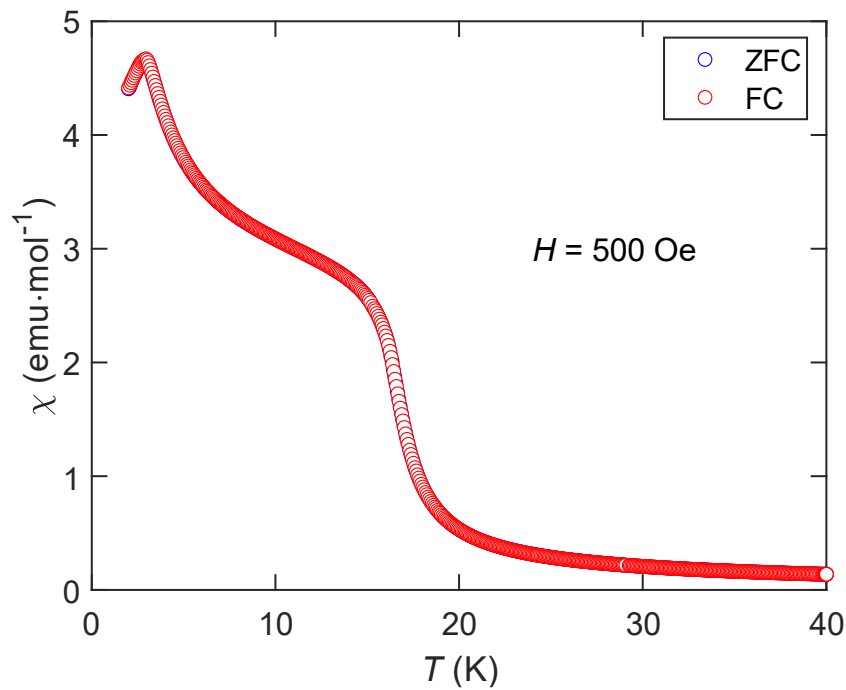

**Figure S4.** Magnetic susceptibility. Magnetic susceptibility of a 15.5(1) mg crystal measured on an MPMS with $H$ = 500 Oe. We attribute the feature at $T$ = 16 K, beyond the transition at $T_N$, to an EuS impurity whose phase fraction varies by sample. The feature appears in measured single crystals but is not visible in the powder sample measured in Figure 1c. Measurements were performed on warming. The molar susceptibility units assume pure EuPd$_3$S$_4$. Uncertainty in the measured mass gives a 0.6% systematic uncertainty.

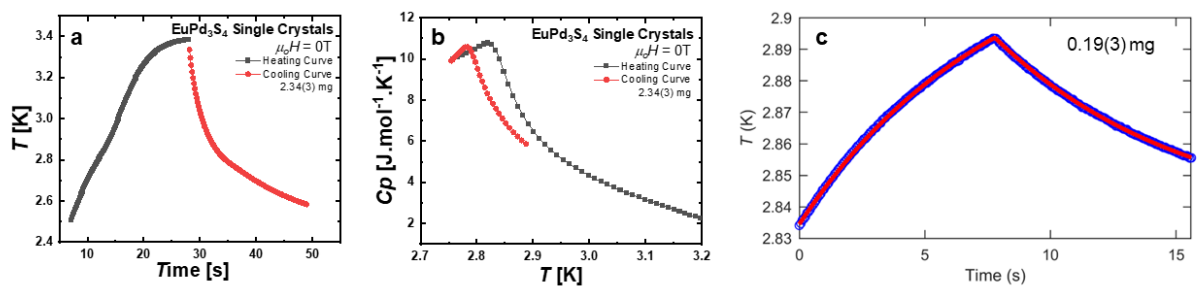

**Figure S5.** Long-pulse heat capacity. (a) Temperature versus time and extracted (b) heat capacity versus temperature data for an EuPd$_3$S$_4$ from long-pulse measurements. There is no plateau in the temperature versus time graph that would indicate a latent heat (and thus first order transition). We attribute the hysteresis in the long pulse measurements to thermal decoupling of the sample from the stage, when its specific heat is larger near $T_N$. (c) This was further confirmed by measurements on smaller samples where excellent fits are obtained to the time dependent sample temperature data acquired during conventional adiabatic heat pulse measurements.

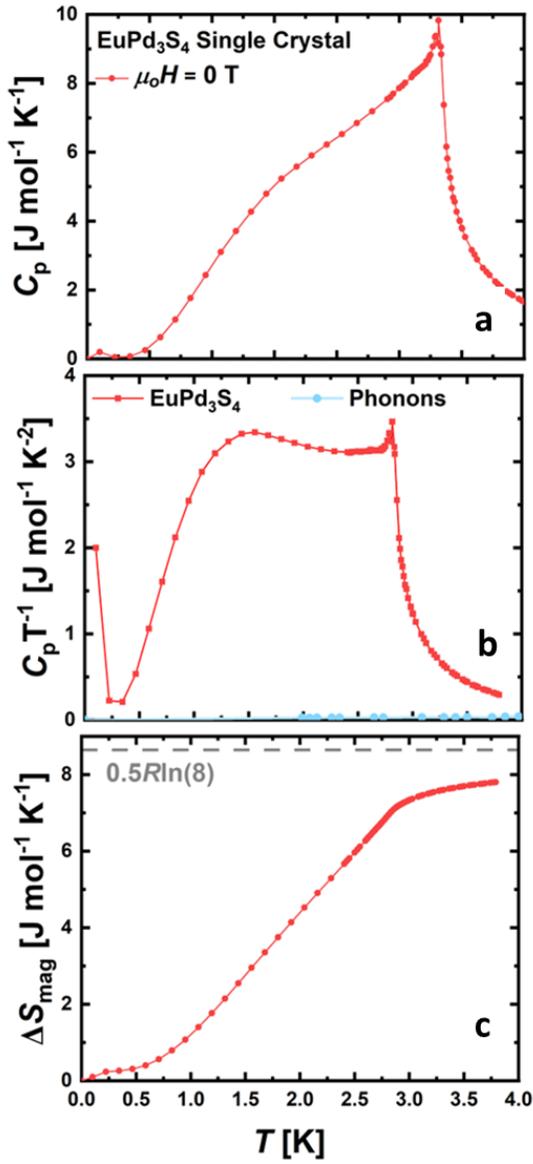

**Figure S6.** Magnetic heat capacity and entropy in EuPd$_3$S$_4$. (a,b) The magnetic heat capacity as determined by subtracting the phonon portion obtained from the reported heat capacity[1] of LaPd$_3$S$_4$ s an estimator of the phonon contribution. (c) The estimated change in magnetic entropy obtained by integration of the data in (b). A EuS impurity likely accounts for the entropy remaining beyond $T_N$.

Heat Capacity Phonon Subtraction

The phonon contributions are subtracted from EuPd$_3$S$_4$ using the heat capacity data from LaPd$_3$S$_4$ through a normalization procedure to account for their molar mass differences [Equation 1].[2]

$$\frac{\Theta^3_{L_mY_sZ_p}}{\Theta^3_{X_mY_sZ_p}} = \frac{mM_X^{1.5} + sM_Y^{1.5} + pM_Z^{1.5}}{mM_L^{1.5} + sM_Y^{1.5} + pM_Z^{1.5}} \quad [1]$$

$\Theta$ is the Debye temperature, $M$ is the mass of the atom, and $(m + s + p)$ is the number of atoms per formula unit.

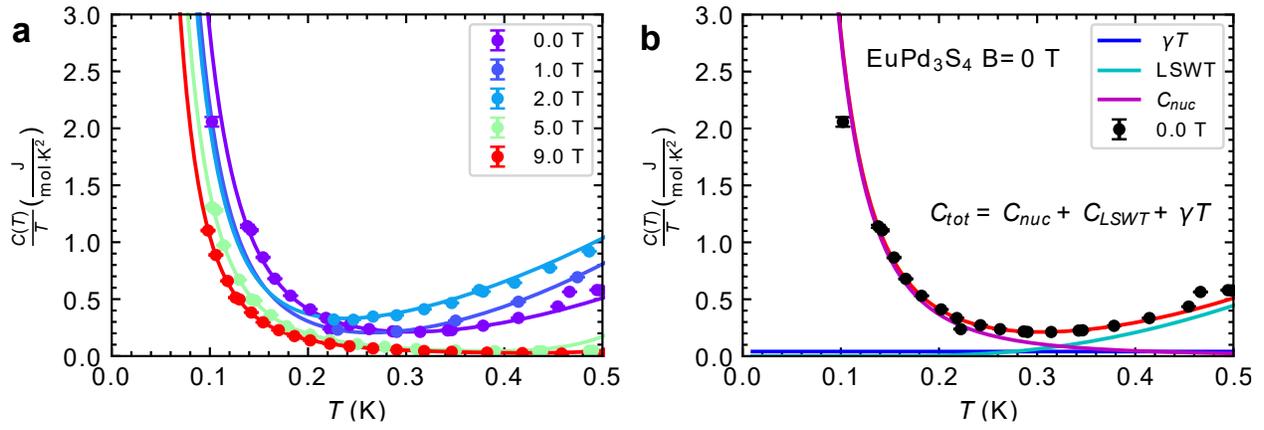

**Figure S7.** Fitting of low temperature upturn to a nuclear Schottky anomaly. (a) Magnetic field dependent heat capacity. (b) Analysis of low temperature upturn as a Schottky anomaly arising from hyperfine enhanced nuclear spin polarization in the presence of Eu magnetic order.

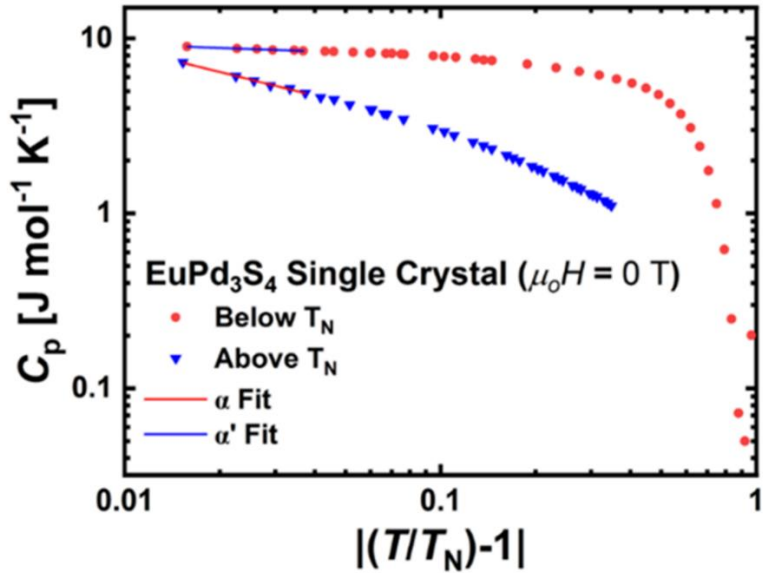

**Figure S8.** Critical scaling analysis of magnetic heat capacity. Double logarithmic plot of specific heat versus reduced temperature $\tau = |T/T_N - 1|$ where $T_N$ = 2.81(1) K was taken to be the datapoint with the largest specific heat. The approximately linear behavior near the maximum in the specific heat yields exponents α = 0.448(14) for $T > T_N$ and α′ = 0.006(7) for $T < T_N$. The different values obtained here indicate that the critical exponent is negative for this phase transition and leads to the alternate analysis described in the following figures S9 and S10.[3] The phonon contribution to the specific heat was subtracted though it represents less than 1.53% of the specific heat capacity in this temperature range (Figure S6).

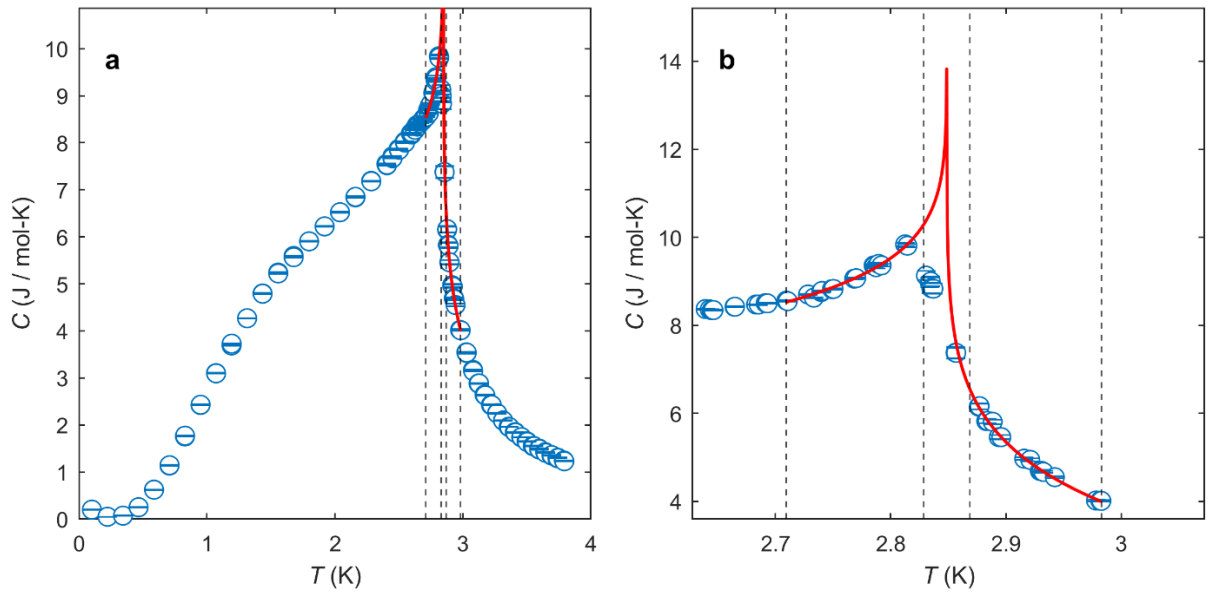

**Figure S9.** Heat capacity critical exponent direct fits to Equation 2. (a,b) Fitting to the pure-power-law model in[3,4] with 0.007<|τ|<0.05 gives a critical exponent of $\alpha$ = -0.11(3). Dashed black lines indicate the two intermediate regions included in our fit. As described in the supplementary methods, the fitting is rather sensitive to the excluded range and initial parameters. But we do find good fits for $\alpha = \alpha'$ < 0. Near $T_N$, the adiabatic heat pulse results in a typical 0.06 K temperature rise during each measurement.

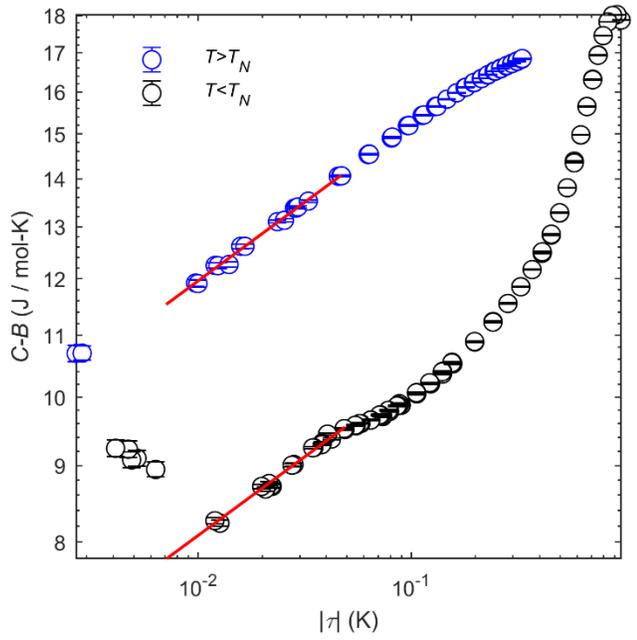

**Figure S10.** Heat capacity critical exponent log-log plots. $C$ is the overall heat capacity, $B$ is a constant term in heat capacity, and $\tau$ is the reduced temperature. When $\alpha$ is negative we expect log($C$-$B$) versus log($\tau$) to be linear near $T_N$. The red line indicates the critical exponent fit for 0.007<|τ|<0.05.

Heat Capacity Critical Exponent Fits

Initially assuming $\alpha > 0$ and a divergent critical contribution $C_p \propto |\tau|^{-\alpha}$ to the specific heat capacity, a linear regression was performed on a double-logarithmic plot of $C_p$ versus $|\tau|$.[5] Data for $|\tau| < 0.05$ were included in the analysis. This approach however, yields different values for $\alpha > 0$ above and below $T_N$, which is inconsistent with the assumed power law form (Figure S8). We therefore attempted the power-law analysis described in [3,4] to see if a reasonable fit with identical $\alpha < 0$ on both sides of the transition can be found (Figure S9). When $\alpha < 0$, there is no longer a diverging power law contribution near $T_N$ and additional terms in the specific heat become significant. The specific form used in the analysis is

$$C = \begin{cases} \dfrac{A}{\alpha}|\tau|^{-\alpha} + B + E\tau, & \tau > 0 \\ \dfrac{A'}{\alpha}|\tau|^{-\alpha} + B + E\tau, & \tau < 0 \end{cases} \quad [2]$$

We expect and observe some rounding near the critical temperature, so the fit was limited to the two temperature ranges $\tau_{min}<|\tau|<\tau_{max}$ about the fitted $T_N$. The error bars reported below were taken to be half the range of the parameter over which the reduced chi-squared remains within the threshold best fit reduced chi-squared. This threshold is given by $\left(1+\frac{1}{\nu}\right)\chi_r^2$ where $\nu$ is the degrees of freedom, taken to be the number of datapoints minus the number of fitted parameters. Fits for $A$, $A'$, $B$, $E$, $\alpha$, and $T_N$ were initially performed using MATLAB's GlobalSearch function,[6] while fmincon alone was used as the local fitting function for error bar calculations. As an example, we find for $0.007<|\tau|<0.05$ the following best fit:

| $A$ (J/mol-K) | $A'$ (J/mol-K) | $B$ (J/mol-K) | $E$ (J/mol-K$^2$) | $\alpha$ | $T_N$ (K) |
| --- | --- | --- | --- | --- | --- |
| 2.0(3) | 1.38(9) | 18(3) | ~0 | -0.11(3) | 2.85(2) |

The experimental uncertainty in the critical temperature is dominated by the 20 mK calibration accuracy of the thermometer. $E$ refined to approximately 0, hitting the lower bound set in the initial round of fitting. The fitting is rather sensitive to the selection of starting parameters and fitting range. We notice that a smaller value of $T_N$ appears to give decent fits with $\alpha = -0.24$ and $\chi_r^2 = 4.3$, though increasing $\alpha$ improves the goodness of fit. For the $0.007<|\tau|<0.05$ range, the best fit with $\alpha = -0.11(3)$ and $\chi_r^2 = 3.5$ corresponds to a universal parameter $P = (1-A/A')/\alpha = 5(1)$. This is consistent with the range of this parameter in previous experimental and theoretical work.[3,4] On the other hand, changing the fitting range to $0.01<|\tau|<0.05$ and allowing the model to refine gives a best fit $\alpha = -0.06(4)$ with $\chi_r^2 = 3.4$. Given the dependence of $\alpha$ on the choice of fitting range, we estimate $\alpha = [-0.3, -0.02]$. For reference, the exponents for the 3D Heisenberg and mean field classes are $\alpha = -0.12(1)$ and $\alpha = 0$, respectively.[7] The 3D Heisenberg class is consistent with short range isotropic magnetic interactions but does not capture longer range dipole interactions. While we don't find a precise value of $\alpha$, we do find $\alpha = \alpha' < 0$ is a reasonable model to our data.

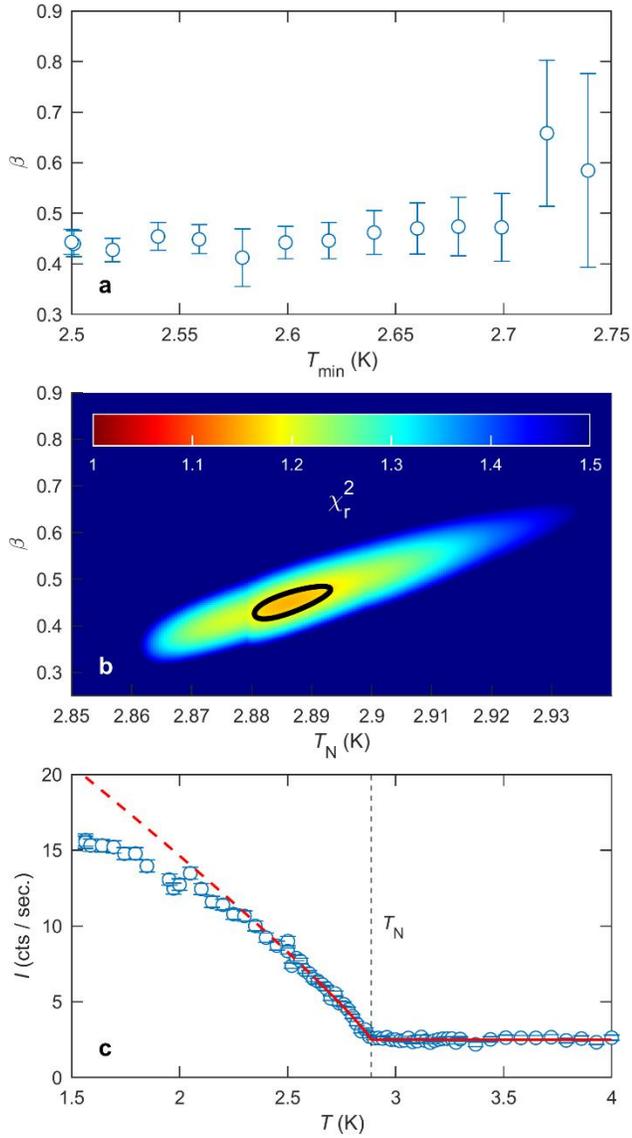

**Figure S11.** Critical scaling analysis of magnetic neutron Bragg diffraction at (½,½,½). After (a) considering several low temperature limits for the fits, and (b) fitting across a grid of $\beta$ and $T_N$ values with $T$ = 2.6 K selected as the low temperature limit, we obtain $\beta$ = 0.45(4) and $T_N$ = 2.89(1) K. The fit is shown in (c) as a solid red line that extends as a dashed line beyond the temperature regime that was included in the analysis. For comparison the exponent for the 3D Heisenberg model is $\beta$ = 0.366(2). The solid black line in (b) indicates the confidence regime derived from a reduced $\chi^2$ analysis. Sample inhomogeneities can give neutron diffraction critical exponents approaching the mean field result. Alternatively, longer range interactions in this metallic system could give rise to mean field like criticality in the accessible regime of reduced temperature. (c) Data from BT7 at the NCNR.

Neutron Diffraction Critical Exponent Fits

We used the temperature dependence of the intensity of the (½,½,½) magnetic Bragg peak as a measure of the squared magnetic order parameter. Only data within the critical regime can be expected to reflect the critical properties. To determine an appropriate lower temperature limit for critical exponent fitting, we carried out fits across several temperature ranges, varying from within 5% to within 15% of $T_N$ and examined the variation in $\beta$ with the fitting range (Figure S11a). Each fit was performed to the power law $I = I_0 + A\tau^{2\beta}$ below $T_N$ and $I = I_0$ beyond $T_N$, where $\tau$ is the reduced temperature $\tau = 1 - \frac{T}{T_N}$ and we fit the parameters $I_0, A, T_N$ and $\beta$. On the one hand, we should fit near enough to the critical transition such that the fitted $\beta$ doesn't deviate much from its value in the critical regime. On the other hand, we should include as many points in the fit as possible. Based on the above compromise, we selected $T$ = 2.6 K as the minimum temperature of our fitting corresponding to staying within 10% of $T_N$. We then proceeded to make power law fits across a grid of candidate $T_N$'s and $\beta$'s varying $I_0$ and $A$ for each point in the $(T_N, \beta)$ plane. Figure S11b shows a color plot of the reduced chi-squares used to identify the best fit values of $\beta$ and $T_N$. The error bars were taken to be half the range of the parameter over which the reduced chi-squared remains within the threshold reduced chi-squared corresponding its best fit value. This is given by $\left(1 + \frac{1}{\nu}\right)\chi_r^2$ where $\nu$ is the degrees of freedom, taken to be the number of datapoints minus the number of fitted parameters. We find that the error in temperature (including the fitted $T_N$) is dominated by the error bar on the resistive thermometer calibration, which is reported to be 0.01 K for our Orange ILL cryostat used in the experiment without a magnet.[8,9] This error does not, however, affect the recorded value of our critical exponent.

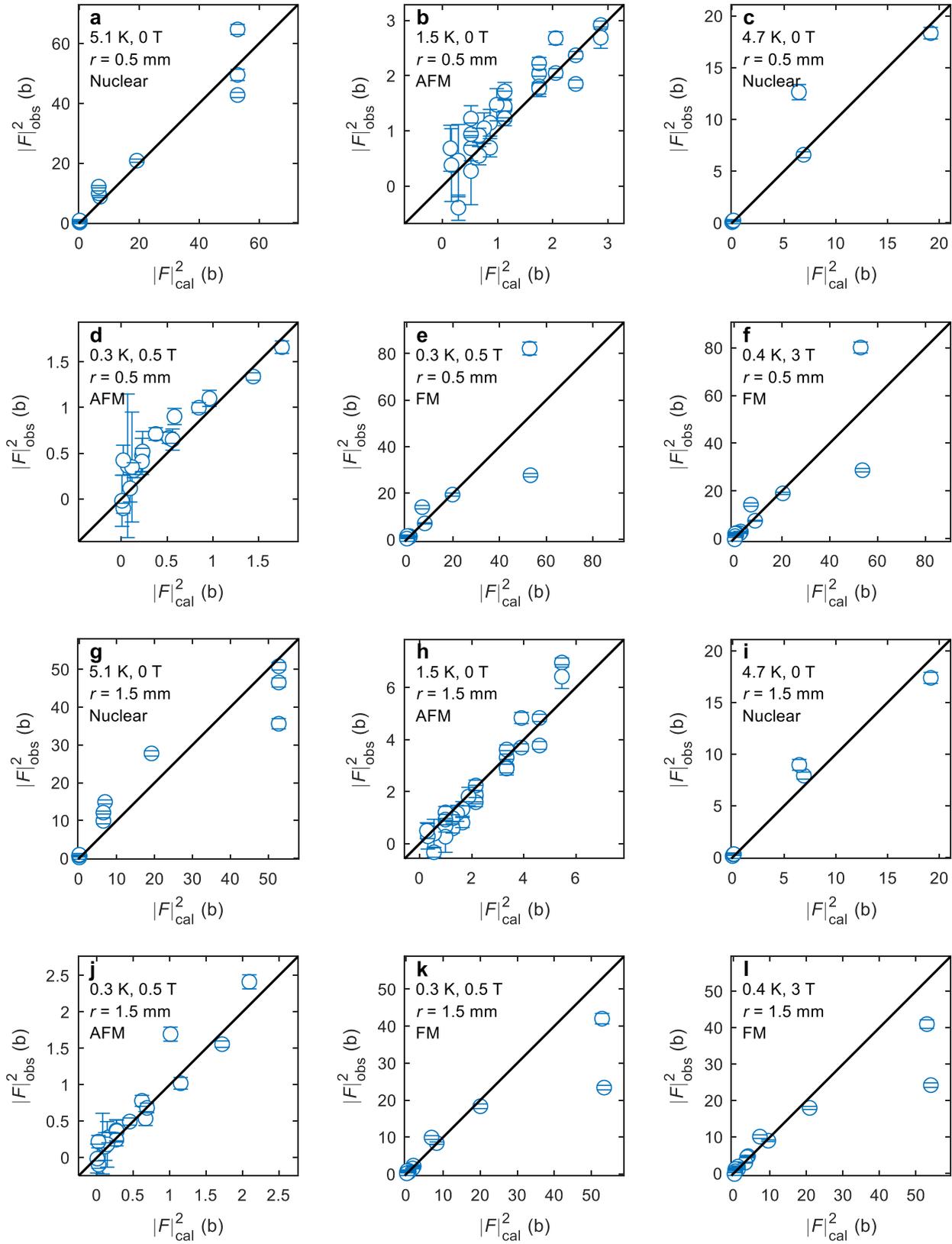

**Figure S12.** Refinement of single crystal neutron diffraction data for EuPd$_3$S$_4$. The plots compare measured and calculated squared structure factors in units of barn for nuclear and magnetic Bragg peaks under distinct thermodynamic conditions. Spherical absorption corrections were employed with a sample radius $r$ = 0.5 mm in frames (a-f), and $r$ = 1.5 mm in (g-l). The sample had an irregular shape with full thickness within the scattering plane ranging from 1 mm to 3 mm. Data in frames (a,b,g,h), were acquired with an ILL-type cryostat while the remaining data were obtained using a He-3 insert within a cryo-magnet. All data were acquired on BT7 at the NCNR.

Neutron Penetration Depth

The absorption coefficient $\mu_a$ of an absorbing compound is given by $\mu_a = \rho\sigma_a$, where $\rho$ is the number density of absorbing atoms in the lattice and $\sigma_a$ is the corresponding absorption cross section. With 25.3 meV neutrons, the absorptions cross sections for naturally occurring Eu, Pd and S are 4530(40) b, 6.9 b and 0.53 b respectively.[10] For our neutron diffraction experiments $E_i = E_f$ = 35 meV so, taking Eu as the only absorbing atom and the absorption cross section to increase linearly with wavelength, $\sigma_a(35\text{ meV}) = 3850$ b.[11,12] The volume of an EuPd$_3$S$_4$ unit cell is 296 Å$^3$ and there are two Eu atoms per cell so $\rho$ = 6.76E-3 atoms/Å$^3$ and $\mu_a$ = 2.60 mm$^{-1}$. The penetration depth for our neutrons is thus only $d$ = 0.384 mm; after traveling through 1 mm of the sample about 92.6% of the incident neutrons have been absorbed. The diameters of the BT7 and EIGER samples were about 2 mm and 1 mm, respectively. The absorption limits the quantity and statistical quality of the neutron diffraction data. It also makes the Bragg intensities sensitive to sample geometry.

Structure Factors

Fits to scans of the observed Bragg peaks were performed in MATLAB. A Gaussian peak with a flat background was obtained by minimizing the reduced chi-squared $\chi_r^2 = \sum_i (I_{\text{obs},i} - I_{\text{calc},i})^2/\sigma_i^2/\nu$, $\nu = N_{\text{obs}} - N_{\text{fit}}$, with intensity given by the equation $I(I_0, A, \theta_0, \sigma) = I_0 + \frac{A}{\sqrt{2\pi}\sigma} e^{-\frac{(\theta-\theta_0)^2}{2\sigma^2}}$ where $I_0$ is a flat background, $A$ is the integrated intensity, $\theta_0$ is the center a3 motor angle, and $\sigma$ is the standard deviation of the Gaussian peak. This was done using the MATLAB function fminsearch. First the width $\sigma$ was determined by fitting to the most prominent peaks for each set of measurements, then a second round of fits with the widths fixed determined $I_0$, $A$, and $\theta_0$. An upper limit of the goodness of fit was obtained for a $\chi_r^2$ distribution from the best fit. The integrated intensity was varied across a range of values about the best fit while the other parameters were fit until the upper $\chi_r^2$ threshold was exceeded. We set the error bar on the fitted integrated intensity equal to the difference between the maximum and minimum values within said limit. The squared structure factors were then calculated from the integrated intensity following a correction for the instrumental resolution using ResLib with a Cooper-Nathans approximation.[13] The corresponding equation is $|F|^2 = QA\sqrt{M_{22}/(2\pi)}/R_0$, where $R_0$ is the normalization prefactor and $M$ is the 4 × 4 resolution matrix.

Absorption Correction

Refinements were done with and without absorption corrections. Absorption corrections were performed in MATLAB by applying an attenuation factor to the squared structure factors. This attenuation factor was obtained following the discussion for a spherical geometry by Maslen.[14] The integration was carried out in MATLAB.

Domain Averaging

For neutron scattering, the structure factor of an antiferromagnetically ordered lattice is affected by the direction of the spins due to a dipolar polarization factor of $\left[1 - (\hat{Q} \cdot \hat{\eta})^2\right]_{\text{av}}$ [15 Eq. 12.30]. Here, $\hat{Q}$ is the unit vector for the observed reflection $(HKL)$ and $\hat{\eta} = (x, y, z) = (\sin\theta \cos\varphi, \sin\theta \sin\varphi, \cos\theta)$ is the unit vector along the uniaxial spin direction within a domain. Space group 200 ($Pm\bar{3}$) has $N$ = 24 crystallographic symmetry operations, 12 of which are related by inversion symmetry.[16,17] But inversion does not change the direction of the magnetic moment (which is a pseudovector), thus only 12 operations must be included in a calculation of $(\hat{Q} \cdot \hat{\eta})^2_{\text{av}}$. It can readily be shown that the cubic domain average of the polarization factor is independent of the uniaxial spin direction within a domain:

$$\left[1 - (\hat{Q} \cdot \hat{\eta})^2\right]_{\text{av}} = 1 - \frac{1}{N}\sum_i^N (\hat{Q} \cdot \hat{\eta}_i)^2 = \frac{2}{3}$$

In other words, $\left[1 - (\hat{Q} \cdot \hat{\eta})^2\right]_{av}$ is independent of $\theta$ and $\varphi$. This means that the direction of the moments within a domain cannot be directly determined by a refinement of the structure against observed intensities for a multi-domain sample. As anticipated by Lovesey, we are thus able to determine the "configuration" of the magnetic lattice rather than the "orientation." The result of this calculation is unchanged for space group 223 ($Pm\bar{3}n$).

Refinements

Single-crystal refinements for the magnetic structure were performed in FullProf against .int datafiles with the observed structure factors. This refinement could be improved by a more accurate absorption correction using detailed knowledge of the sample geometry, by machining the sample into a spherical or cylindrical shape, and by studying an isotopically substituted sample. For 0-field measurements, powder diffraction from a thin annular sample is also a possibility. For 0.5 T antiferromagnetic refinement, we would expect the metamagnetic transition to produce a magnetic phase with the staggered magnetization perpendicular to the [1-10] applied field. We find that the fit with moments along the [110] direction ($\chi^2$ = 5 for $r$ = 0.5 mm) converged significantly better than for [001] oriented spins ($\chi^2$ = 127 for $r$ = 0.5 mm). The observed squared structure factors plotted against calculated squared structure factors are in Figure S12.

BCC versus SC moment sizes

While the Eu atoms are arranged on a bcc lattice, only $Eu^{2+}$ is magnetic. One may consider two configurations of the magnetic moments in $EuPd_3S_4$. In the first model, all magnetic ions are distributed with equal probability in space, so the corner and center sites are each 50% magnetic and 50% non-magnetic. In the second model, each ion takes a different position in the unit cell so one simple cubic sublattice is 100% magnetic while the other is entirely non-magnetic. Now we field polarize all of the magnetic ions by applying a magnetic field in excess of 3 T. How can we distinguish between these two models from neutron diffraction? Because all sites are equally polarized along the same axis, the first model will only enhance nuclear peaks. The magnetic peaks will continue to obey reflections conditions of the crystallographic space group (223). Among these conditions is that a reflection of the form (HHL) must have L even. On the other hand, the second model gives a polarized simple cubic structure, which breaks the n-glide symmetry relating the corner and center sites. By breaking this symmetry, the aforementioned reflection condition no longer applies and new integer-valued magnetic reflections appear. This is how we understand the presence of the {100} type reflection in our diffraction data. Additionally, there is an important difference between the moment size required to produce reflections that are common to both models. This is important even in the antiferromagnetic state. The bcc model has twice the number of occupied sites as the sc model. But half of these sites are non-magnetic so, on average, the moment size is half as large and, for a given moment, both models will have the same saturation magnetization. These factors of two do not cancel in the diffraction results, however, because intensity varies in proportion to the squared magnetization. For the disordered body-centered-cubic model to give the same observed intensity as the charge ordered simple cubic lattice model, the moment must be increased by a factor of $\sqrt{2}$. With sufficient experimental accuracy, the size of the refined moments can thus be used to exclude one model in favor of the other.

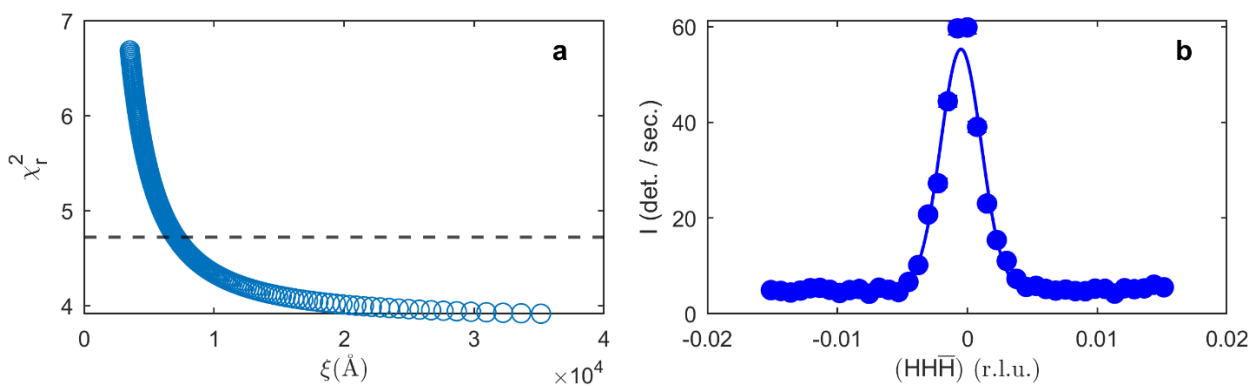

**Figure S13.** (½,½,½) Correlation length. (a) Fitting the (½,½,½) reflection at 1.5 K and 0 T to a Voigt function with Gaussian resolution width fixed to the average value of prominent nuclear peaks in the paramagnetic state. The dashed line indicates a threshold goodness of fit 20% greater than the best fit. Because the magnetic Bragg peak thus is resolution limited, we obtain a lower limit on the correlation length from the inverse Gaussian half width at half maximum: $\xi_{min} = 580(20)$ Å. (b) A Gaussian with the instrumental resolution width is fit to the (½,½,½) magnetic peak, corresponding to true long range order.

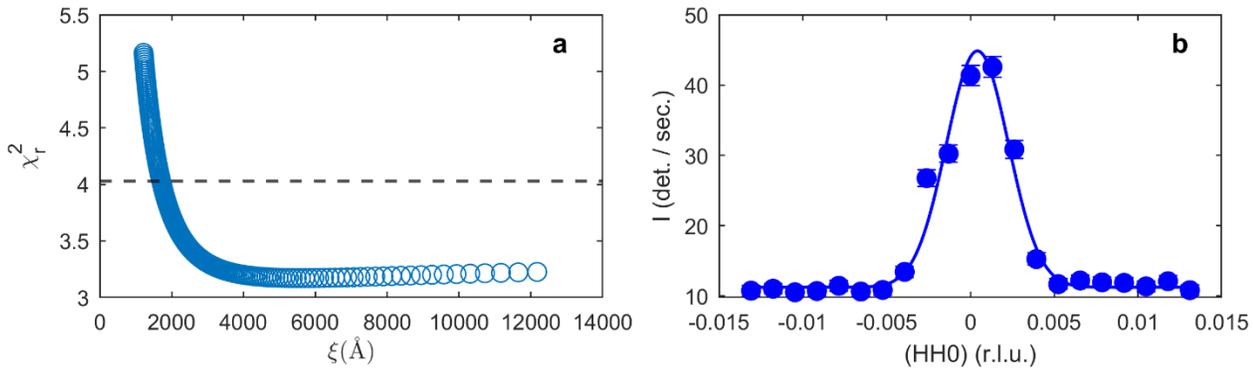

**Figure S14.** (001) Correlation length. (a) Fitting the (001) reflection at 0.4 K and 3 T to a Voigt function with Gaussian width fixed to the average value of prominent nuclear peaks in the paramagnetic phase. A shallow minimum in the goodness of fit with a broad range of large correlation lengths is obtained. The dashed line indicates a threshold goodness of fit 20% greater than the best fit. Because the magnetic Bragg peak is resolution limited, we obtain a lower limit on the correlation length from the inverse Gaussian half width at half maximum: $\xi_{min} = 490(20)$ Å. (b) A Gaussian fit with width fixed to the instrumental resolution is fit to the magnetic (001) peak, corresponding to truly long range order.

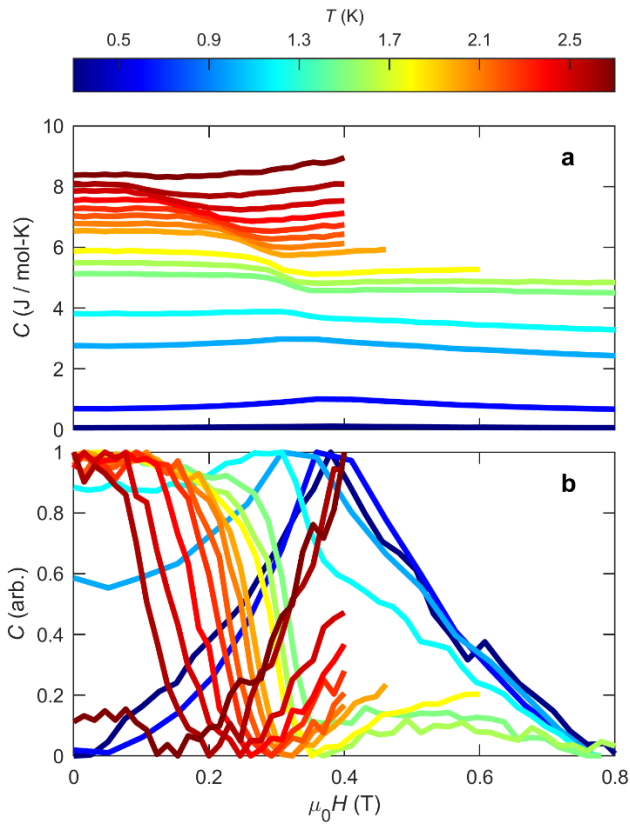

**Figure S15.** Heat capacity data versus applied field indicating a metamagnetic transition. (a) Raw specific heat capacity data versus $\mu_0 H$ for a range of temperatures. The field was applied along the (122) direction. The temperatures measured are indicated by the color bar and were specifically 0.3 K, 0.6 K, 1.0 K, 1.2 K, 1.5 K, 1.6 K, 1.8 K, and 2.0 K to 2.7 K in 0.1 K intervals. (b) Specific heat capacity data versus applied field $\mu_0 H$ normalized so the minimum at each field is 0 while the maximum is 1. For lower

temperatures the putative spin-flop transition is marked by a peak. For temperatures beyond 1.5 K however, only a shoulder remains that moves to lower fields with increasing $T$. The small upper peak moving in from the right for temperatures beyond 2 K marks the transition to a field polarized state. This field driven transition eventually meets the lower field transition at $T_N$ = 2.85(6) K. Because of the small sample mass and high relative error, heat capacity values have been calibrated to the larger 2.34 mg sample's heat capacity by overplotting identical temperature ranges and applying a scale factor.

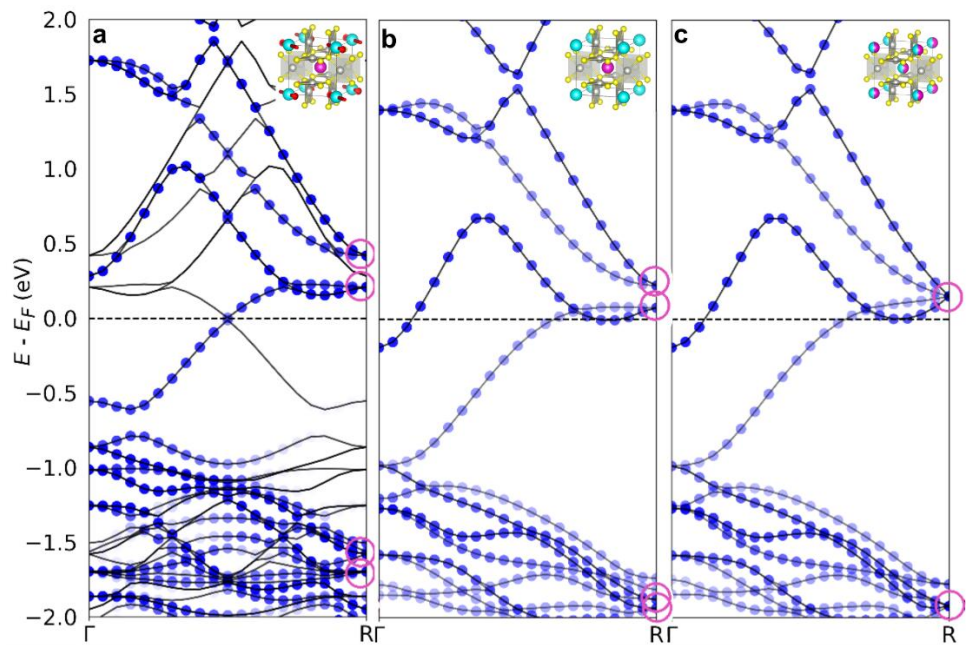

**Figure S16.** DFT Calculations without spin-orbit coupling. (a) Unfolded band structure from the G-type simple cubic AFM magnetic state below $T_N$ = 2.85(6) K showing retention of the 6-fold degeneracies when the magnetic moments lie along an axis with $Fmmm'$ magnetic symmetry. (b) Below 340 K, $Eu^{2+}/Eu^{3+}$ charge order, corresponding to $Pm\bar{3}$ crystal symmetry, splits the 12-fold states into pairs of 6-fold ones, circled (these split into 4-fold and 2-fold degeneracies with addition of SOC). (c) The crystal structure of $EuPd_3S_4$ at $T$ = 370(2) K corresponds to $Pm\bar{3}n$ crystal symmetry, with a single crystallographically unique Eu site; the n glide plane protects 12-fold degenerate states at the **R** point, circled (these split into 8-fold and 4-fold degeneracies with addition of SOC).

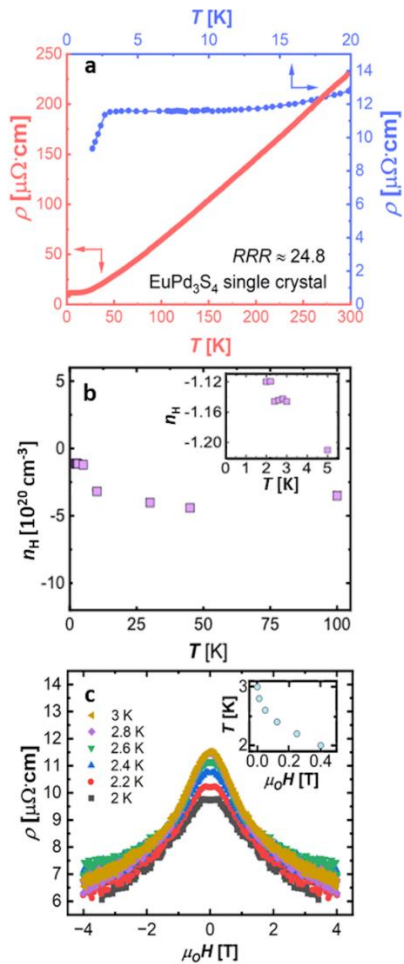

**Figure S17.** Transport behavior in EuPd$_3$S$_4$ single crystals with current perpendicular to the (122) direction. (a) Resistivity as a function of temperature from $T$ = 1.8 K - 300 K at zero applied magnetic field. (b) Carrier concentration, extracted from fits of the linear region of Hall measurements, at a few different temperatures. (c) Field-dependent resistivity measurement at $T$ = 2 K, 2.2 K, 2.4 K, 2.6 K, 2.8 K, and 3 K. The inset shows the temperature for which $\rho$ is at a maximum, as a function of magnetic field applied along the (122) direction.

Electrical transport measurements

Resistivity measurements were carried out in a physical property measurement system (PPMS-14T and PPMS-9T, Quantum Design) using the Resistivity option. Current was applied in the (122) plane, and magnetic fields perpendicular to that along the (122) direction. For longitudinal resistivity, linear contacts were made on the naturally grown crystals by silver paint and 25 μm platinum wires. The longitudinal and Hall resistivities were measured in 4-wire and 5-wire geometry respectively using a current of 3.0 mA–5.0 mA in a temperature range from 0.45 K to 300 K and magnetic fields up to 14 T. The maxima in the field-dependent resistivity were obtained by visual inspection of Figure S17c.

Resistivity

As seen in S17a, the basic resistivity trend in EuPd$_3$S$_4$ is metallic (resistivity increases with temperature) in the absence of an applied magnetic field. The anomaly at $T_N$ = 2.8 K is associated with the antiferromagnetic ordering in EuPd$_3$S$_4$. The carrier concentration in S17b is consistent with that of a normal metal across a few temperature ranges. In S17c, the negative magnetoresistance in EuPd$_3$S$_4$ may be associated in part with a reduction in spin disorder scattering as the material is magnetized. The anomalies observed below the $T_N$ in S17c are consistent with $R(T)$ in S17a. The anomaly trends are further explained in S17c inset with a monotonic decrease close to $T_N$.

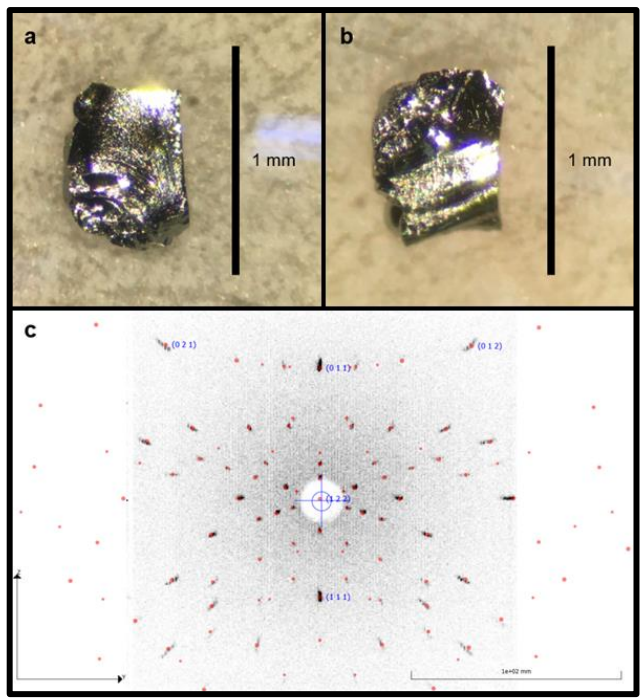

**Figure S18.** Crystals and Laue diffraction. (a,b) Both sides of the platelike 0.19(3) mg EuPd$_3$S$_4$ single crystal used in the heat capacity measurement reported in Figures 2 and 3. (c) Laue backscattering pattern showing alignment of the applied *B*-field (perpendicular to the plane) along (122). Simulation performed in QLaue.[18]

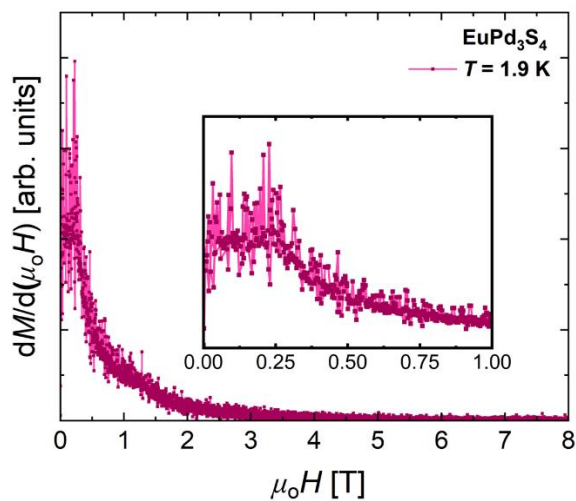

**Figure S19.** Field derivative of the magnetization along (122). An anomaly near 0.25 T relates to the anomaly in specific heat and may be a high temperature manifestation of the spin flop observed in neutron diffraction at lower temperatures.

## Supplementary Tables

(a)

| Formula | EuPd$_3$S$_4$ |
|---|---|
| Sample Identifier | Crystal A |
| Crystal system | Cubic |
| Space group | $Pm\bar{3}$ (No. 200) |
| Temperature (K) | 213(2) K |
| Z | 2 |
| F.W. (g/mol) | 599.40 |
| $a$ (Å) | 6.67570(7) |
| $V$ (Å$^3$) | 297.502(1) |
| Radiation | Mo Kα, λ = 0.71073 Å |
| Reflections | 7189 |
| Unique reflections | 269 |
| Refined parameters | 11 |
| Goodness of fit | 1.124 |
| $R(F)$[a] | 0.0413 |
| $R_w(|F_o|^2)$[b] | 0.0910 |

[a] $R(F) = \Sigma||F_o| - |F_c||/ \Sigma|F_o|$

[b] $R_w(|F_o|^2) = [\Sigma w(|F_o|^2-|F_c|^2)/ \Sigma w|F_o|^2]^{1/2}$

(b)

| Atom | Wyckoff Positions | Occ. | x | y | z | $U_{eq}$ (Å$^2$) |
|---|---|---|---|---|---|---|
| Eu1 | 1$a$ | 1 | 0 | 0 | 0 | 0.0070(6) |
| Eu2 | 1$b$ | 1 | ½ | ½ | ½ | 0.0066(6) |
| Pd | 6$f$ | 1 | 0.24787(1) | 0 | ½ | 0.0070(9) |
| S | 8$i$ | 1 | 0.24614(1) | 0.24614(1) | 0.24614(1) | 0.0070(7) |

(c)

| | $U(1,1)$ (Å$^2$) | $U(2,2)$ (Å$^2$) | $U(3,3)$ (Å$^2$) | $U(1,2)$ (Å$^2$) | $U(1,3)$ (Å$^2$) | $U(2,3)$ (Å$^2$) |
|---|---|---|---|---|---|---|
| Eu1 | 0.0070(2) | 0.0070(2) | 0.0070(2) | 0 | 0 | 0 |
| Eu2 | 0.0066(2) | 0.0066(2) | 0.0066(2) | 0 | 0 | 0 |
| Pd | 0.00874(3) | 0.00554(3) | 0.00596(3) | 0 | 0 | 0 |
| S | 0.00655(2) | 0.00655(2) | 0.00655(3) | 0.00019(1) | 0.00019(1) | 0.00019(1) |

**Table S1.** Single crystal structure of EuPd$_3$S$_4$ ("crystal A") in $Pm\bar{3}$ at $T$ = 213(2) K. (a) Single crystal refinement parameters. (b) Atomic coordinates, occupancies, and isotropic displacement parameters ($U_{eq}$ is defined as one-third of the trace of the orthogonalized $U_{ij}$ tensor (Å$^2$)). (c) Anisotropic displacement parameters.

(a)

| Formula | EuPd$_3$S$_4$ |
|---|---|
| Sample Identifier | Crystal B |
| Crystal system | Cubic |
| Space group | $Pm\bar{3}$; (No. 200) |
| Temperature (K) | 301(2) |
| Z | 2 |
| F.W. (g/mol) | 599.40 |
| $a$(Å) | 6.6632(2) |
| $V$ (Å$^3$) | 295.83(3) |
| Radiation | Mo Kα, λ = 0.71073 Å |
| θ range (°) | 3.057 – 33.127 |

| | | |
|---|---|---|
| Reflections; $R_{int}$ | | 1701; 0.0209 |
| Unique reflections | | 232 |
| Refined parameters | | 11 |
| Goodness of fit | | 1.439 |
| $R(F)$ | | 0.0255 |
| $R_w(|F_o|^2)$ | | 0.0534 |
| Diffraction peak and hole (e⁻/ Å³) | | 2.210; -3.375 |

(b)

| Atom | Wyckoff Position | Occ. | x | y | z | $U_{eq}$ (Å²) |
|---|---|---|---|---|---|---|
| Eu1 | 1a | 1 | 0 | 0 | 0 | 0.0072(3) |
| Eu2 | 1b | 1 | ½ | ½ | ½ | 0.0080(3) |
| Pd  | 6f | 1 | 0.24680(9) | 0 | ½ | 0.0072(2) |
| S   | 8i | 1 | 0.24578(13) | 0.24578(13) | 0.24578(13) | 0.048(11) |

**Table S2.** Single crystal structure of EuPd₃S₄ ("crystal B") in $Pm\bar{3}$ at $T$ = 301(2) K. (a) Single crystal refinement parameters. (b) Atomic coordinates, occupancies, and isotropic displacement parameters ($U_{eq}$ is defined as one-third of the trace of the orthogonalized $U_{ij}$ tensor (Å²)).

(a)

| | |
|---|---|
| Formula | EuPd₃S₄ |
| Sample Identifier | Crystal B |
| Crystal system | Cubic |
| Space group | $Pm\bar{3}n$; (No. 223) |
| Temperature (K) | 370(2) |
| Z | 2 |
| F.W. (g/mol) | 599.40 |
| $a$(Å) | 6.6738(2) |
| $V$ (Å³) | 297.25(3) |
| Radiation | Mo Kα, λ = 0.71073 Å |
| θ range (º) | 4.319 – 38.053 |
| Reflections; $R_{int}$ | 1639; 0.0401 |
| Unique reflections | 160 |
| Refined parameters | 7 |
| Goodness of fit | 1.422 |
| $R(F)$ | 0.0428 |
| $R_w(|F_o|^2)$ | 0.1071 |
| Diffraction peak and hole (e⁻/ Å³) | 1.850; -2.089 |

(b)

| Atom | Wyckoff Position | Occ. | x | y | z | $U_{eq}$ (Å²) |
|---|---|---|---|---|---|---|
| Eu | 2a | 1 | 0 | 0 | 0 | 0.0092(5) |
| Pd | 6c | 1 | ¼ | 0 | ½ | 0.0091(4) |
| S  | 8e | 1 | ¼ | ¼ | ¼ | 0.0085(6) |

**Table S3.** Single crystal structure of EuPd₃S₄ ("crystal B") in $Pm\bar{3}n$ at $T$ = 370(2) K. (a) Single crystal refinement parameters. (b) Atomic coordinates, occupancies, and isotropic displacement parameters ($U_{eq}$ is defined as one-third of the trace of the orthogonalized $U_{ij}$ tensor (Å²)).

| Field (T) | Fraction Eu$_2+$ | $\mu_{eff}$ | $\gamma$ (mJ/mol/K) | $\Delta$ (meV) | A (a.u.) |
|---|---|---|---|---|---|
| 0 | 0.50 (fixed) | 8.0(1) | 40(10) | 0.14(2) | 77(10) |
| 1 | 0.50 (fixed) | 7.3(7) | 2(8) | 0.11(2) | 120(20) |
| 2 | 0.50 (fixed) | 7(3) | 90(20) | 0.04(3) | 90(5) |
| 5 | 0.50 (fixed) | 5.8(1) | 12(5) | 0.29(5) | 38(4) |
| 9 | 0.50 (fixed) | 5.0(1) | 0.0(2) | 0.0(1) | 6(1) |

**Table S4.** Parameters from fitting of the nuclear Schottky anomaly. Analysis of low temperature upturn as a Schottky anomaly arising from nuclear spins freezing in the presence of Eu magnetic order.[19,20] Heat capacity below $T = 0.4$ K was fitted by a superposition of three contributions. $C_{\text{nuc}}(T) = \frac{1}{Z k_B T^2}\left[\sum_i E_i^2 e^{\frac{-E_i}{k_B T}} - \frac{1}{Z}\left(\sum_i E_i e^{\frac{-E_i}{k_B T}}\right)^2\right]$ is the specific heat associated with nuclear spin states arising from the hyperfine Hamiltonian $H = a' I_z + P\left(I_z^2 - \frac{1}{3} I(I+1)\right)$ where $a' = a\langle J_z\rangle$ and $\langle J_z\rangle = \frac{\mu_{\text{eff}}}{g_L}$. $C_{\text{elec}}(T) = \gamma T$ is associated with the electronic density of states at the Fermi level. Finally from linear spin wave theory, $C_{\text{LSWT}} \approx \frac{\partial U}{\partial T}$ for a generic 3D magnon is modeled by $U(T) = \int_\Delta^\infty \frac{\epsilon g(\epsilon)}{e^{\beta\epsilon}-1} d\epsilon$ where $\epsilon$ is the spin wave excitation energy, $\Delta$ is the excitation gap, $\beta = \frac{1}{k_B T}$, and $g(\epsilon)$ is the magnon density of states. For a three-dimensional antiferromagnet with dispersion relation $\epsilon(q) = \sqrt{\Delta^2 + (cq)^2}$ where $c$ is the spin wave velocity, the density of states is $g(\epsilon) = \frac{V\eta}{2 c^3 \pi^2}\epsilon\sqrt{\epsilon^2 - \Delta^2}$ where $V$ is the unit cell volume and $\eta$ is a constant. Without information on dispersion, we take the entire prefactor $\frac{V\eta}{2 c^3 \pi^2}$ to be a single constant A. The gap was found to be about 0.1 meV/1.2 K. Because our model works below the gap, we fit to data below $T = 0.4$ K.

| $T$ (K) | $\mu_0 H$ (T) | Structure | $\mu_{0.5\text{mm}}$ ($\mu_B$) | $R_{0.5\text{mm}}$ | $\chi^2_{0.5\text{mm}}$ | $\mu_{1.5\text{mm}}$ ($\mu_B$) | $R_{1.5\text{mm}}$ | $\chi^2_{1.5\text{mm}}$ | $\mu$ ($\mu_B$) |
|---|---|---|---|---|---|---|---|---|---|
| 5.1 | 0 | Nuclear | - | 19 | 45 | - | 27 | 75 | - |
| 1.5 | 0 | AFM | 8.0(3) | 21.7 | 6 | 11.0(5) | 19 | 8 | 9(2) |
| 4.7 | 0 | Nuclear | - | 20 | 13 | - | 17 | 9 | - |
| 0.3 | 0.5 | AFM | 6.1(3) | 31 | 5 | 6.7(4) | 23 | 6 | 6.4(3) |
| 0.3 | 0.5 | FM | 5(2) | 42 | 115 | 6(4) | 43 | 196 | 5.0(5) |
| 0.4 | 3 | Polarized PM | 6(2) | 42 | 120 | 8(3) | 44 | 200 | 7.1(6) |

**Table S5.** Single crystal magnetic refinement with spherical absorption correction. The first two rows report results acquired at $\mu_0 H = 0$ T with the sample in an ILL-type cryostat. The following four rows employed a He-3 insert within a cryomagnet. Refinements were performed using FullProf.[21] Here $R = 100 \sum_n \left[\left||F_{\text{obs},n}|^2 - \sum_k |F_{\text{calc},k}|^2\right|\right] / \sum_n |F_{\text{obs},n}|^2$, where $F$ is the structure factor, $n$ indexes the Bragg peaks, and the k-summation is over the nuclear and magnetic contributions to each of these Bragg peaks. $\chi^2 = M/(N_{\text{obs}} - p)$, where $M = \sum_n w_n \left(|F_{\text{obs},n}|^2 - \sum_k |F_{\text{calc},k}|^2\right)^2$ is the refinement's objective function, $N_{\text{obs}}$ is the total number of observations, $p$ is the number of fitted parameters, and $w_n$ is the weight of an observation i.e. the inverse square of the variance. Error bars on the moment size correspond to a 20% increase in $\chi^2$ beyond the minimum value. Because of strong neutron absorption by Eu and its irregular shape, spherical models with radii $r = 0.5$ mm and $r = 1.5$ mm were compared corresponding to the limits of the effective sample radius so as to obtain an estimate of the systematic error on $\mu$ arising from absorption effects. The final column reports the ordered moment obtained by averaging the results of the refinement with each of the extremal radii associated with the irregular sample shape.